\documentclass[acmtog,]{acmart}

\usepackage[ruled,linesnumbered]{algorithm2e} % For algorithms
\usepackage{multirow} % Add this for multirow support in tables
\usepackage{subfigure}
\usepackage{wrapfig}

\begin{document}

\title{Curly Hair Simulation using Curly Finite Elements}


\author{Xinming Pei}
\authornote{Work done during an internship at Style3D Research.}
\affiliation{%
	\institution{Zhejiang University}
	\city{Hangzhou}
	\country{China}}
\affiliation{%
	\institution{Style3D Research}
	\city{Hangzhou}
	\country{China}}
\email{xmpei@zju.edu.cn}

\author{Zhendong Wang}
\authornote{Corresponding author}
\email{wang.zhendong.619@gmail.com}
\affiliation{%
	\institution{Style3D Research}
	\city{Hangzhou}
	\country{China}}

\author{Lei Lan}
\affiliation{%
	\institution{Zhejiang University}
	\city{Hangzhou}
	\country{China}}
\email{leilan@zju.edu.cn}

\author{Weiwei Xu}
\authornotemark[2]
\affiliation{%
	\institution{Zhejiang University}
	\city{Hangzhou}
	\country{China}}
\email{xww@cad.zju.edu.cn}

\author{Yin Yang}
\affiliation{%
	\institution{University of Utah}
	\city{Utah}
	\country{USA}}
\email{yangzzzy@gmail.com}

\author{Huamin Wang}
\affiliation{%
	\institution{Style3D Research}
	\city{Hangzhou}
	\country{China}}
\email{wanghmin@gmail.com}


\begin{abstract}

    Realistic simulation of curly hair is challenging due to the tight coupling between macroscopic strand deformation and high-frequency geometric details such as waves and helices. In this paper, we propose a curly hair model that decomposes each strand into curly elements, consisting of a rod-base configuration and an analytically defined high-frequency wrinkles represented by planar waves or volumetric helices, with deformation governed primarily by bending for wavy hair and twisting for spiral hair. A curvature-energy splitting scheme separates stretching, buckling, and bending contributions, and efficient energy approximations improve numerical robustness and reduce computational cost without degrading visual fidelity. We further introduce a hybrid collision handling strategy that combines coarse collision proxies for the base configuration with analytical treatment of high-frequency details, and adapt guide-strand interpolation to our curly finite-element representation to generate dense hair while preserving fine structure. Experiments demonstrate stable, efficient, and visually faithful simulation of curly hair across diverse scenarios.
\end{abstract}




% \begin{figure*}
%     \centering
%     \subfigure[caption]{\includegraphics[width=0.16\linewidth]{figures/teaser1.png}}
%     \subfigure[caption]{\includegraphics[width=0.16\linewidth]{figures/teaser2.png}}
%     \subfigure[caption]{\includegraphics[width=0.16\linewidth]{figures/teaser3.png}}
%     \subfigure[caption]{\includegraphics[width=0.16\linewidth]{figures/teaser4.png}}
%     \subfigure[caption]{\includegraphics[width=0.16\linewidth]{figures/teaser5.png}}
%     \subfigure[caption]{\includegraphics[width=0.16\linewidth]{figures/teaser6.png}}

%     \caption{Our wave-based hair simulation method effectively captures high-frequency details such as curls and waves while maintaining computational efficiency.}
%     \label{fig:teaser} 
% \end{figure*}

\begin{teaserfigure}
    \centering
    \subfigure[Base Strands]{\includegraphics[width=0.16\linewidth]{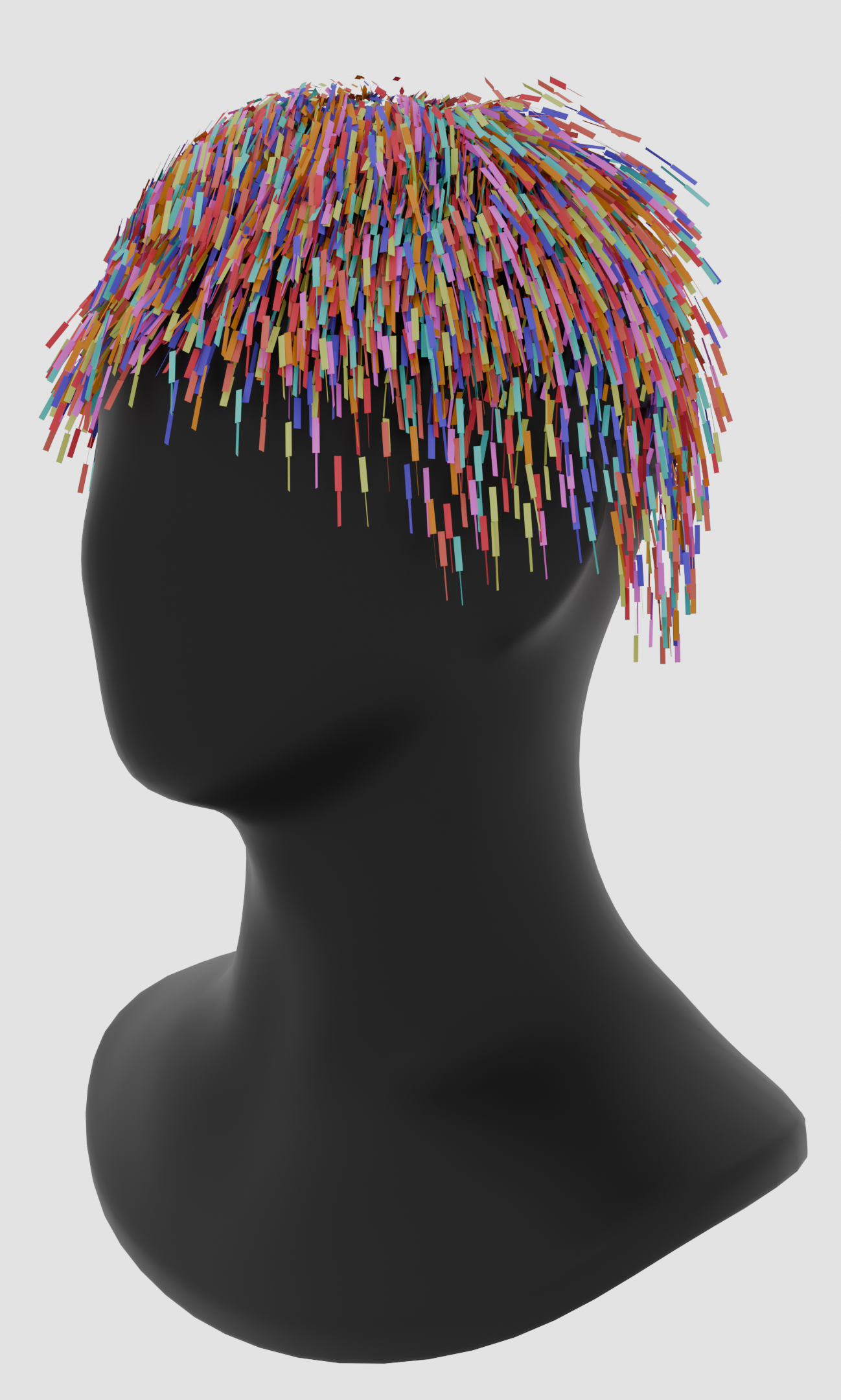}}
    \subfigure[Wavy Strands]{\includegraphics[width=0.16\linewidth]{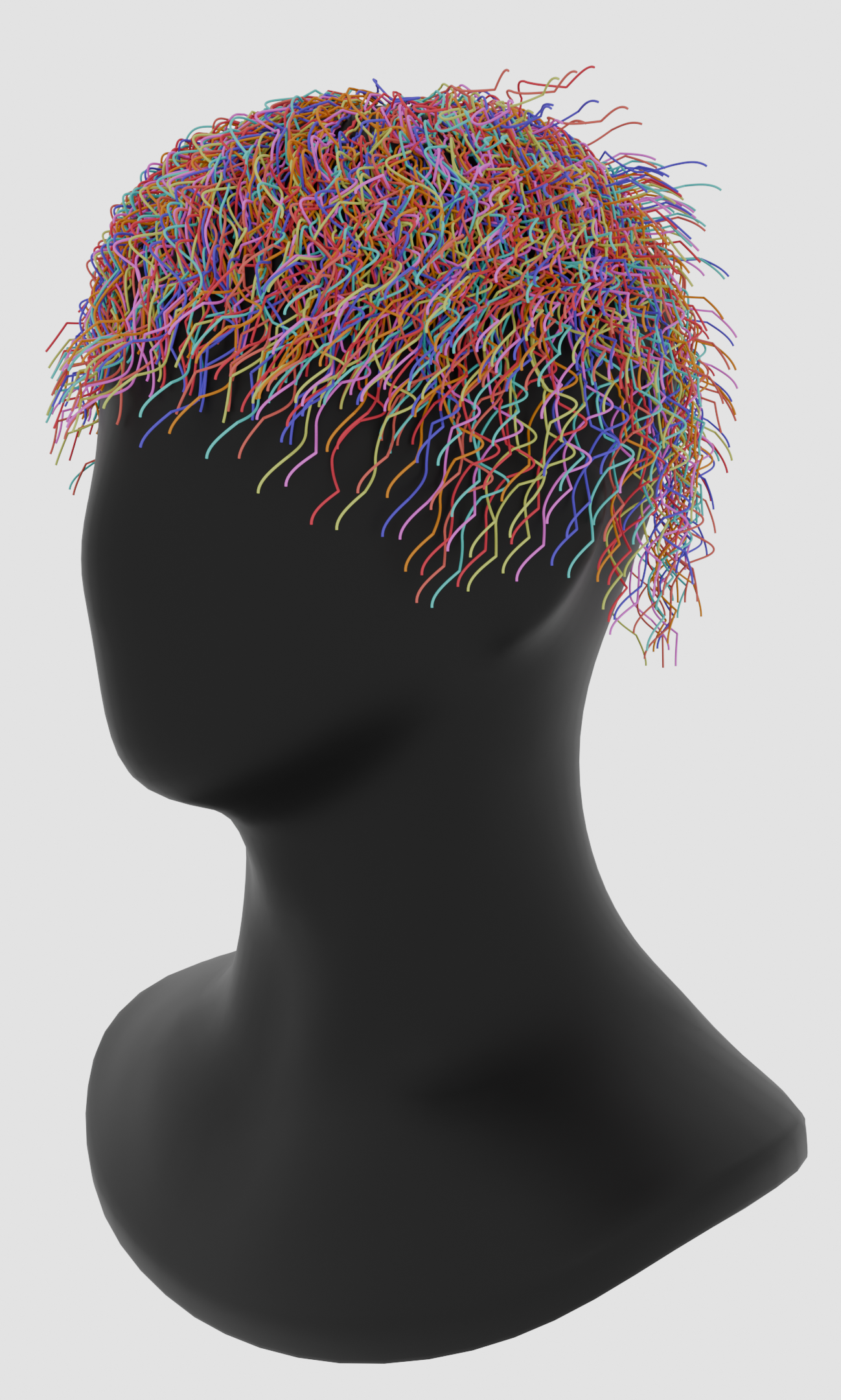}}
    \subfigure[Twisted Perm]{\includegraphics[width=0.16\linewidth]{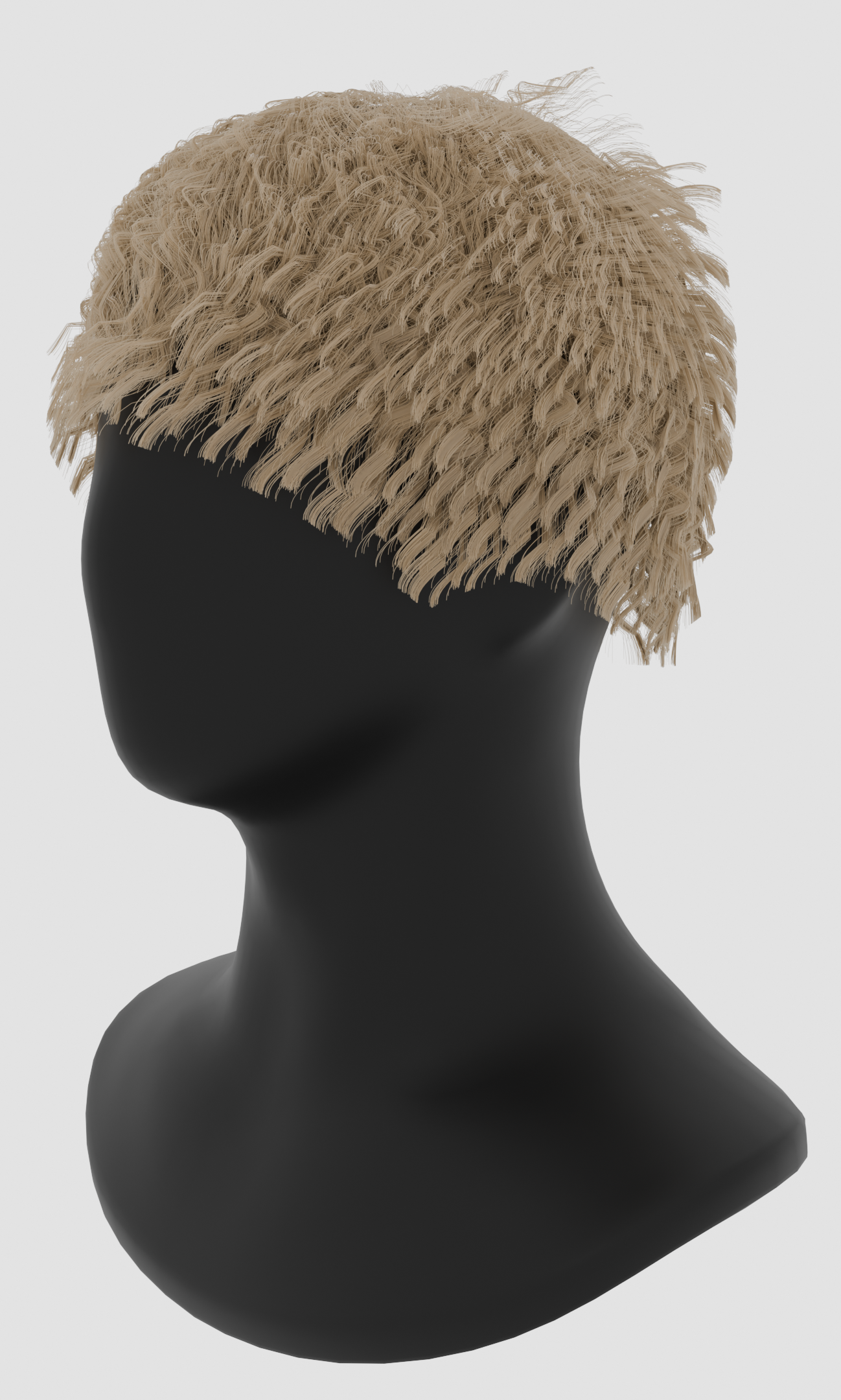}}
    \subfigure[Base Strands]{\includegraphics[width=0.16\linewidth]{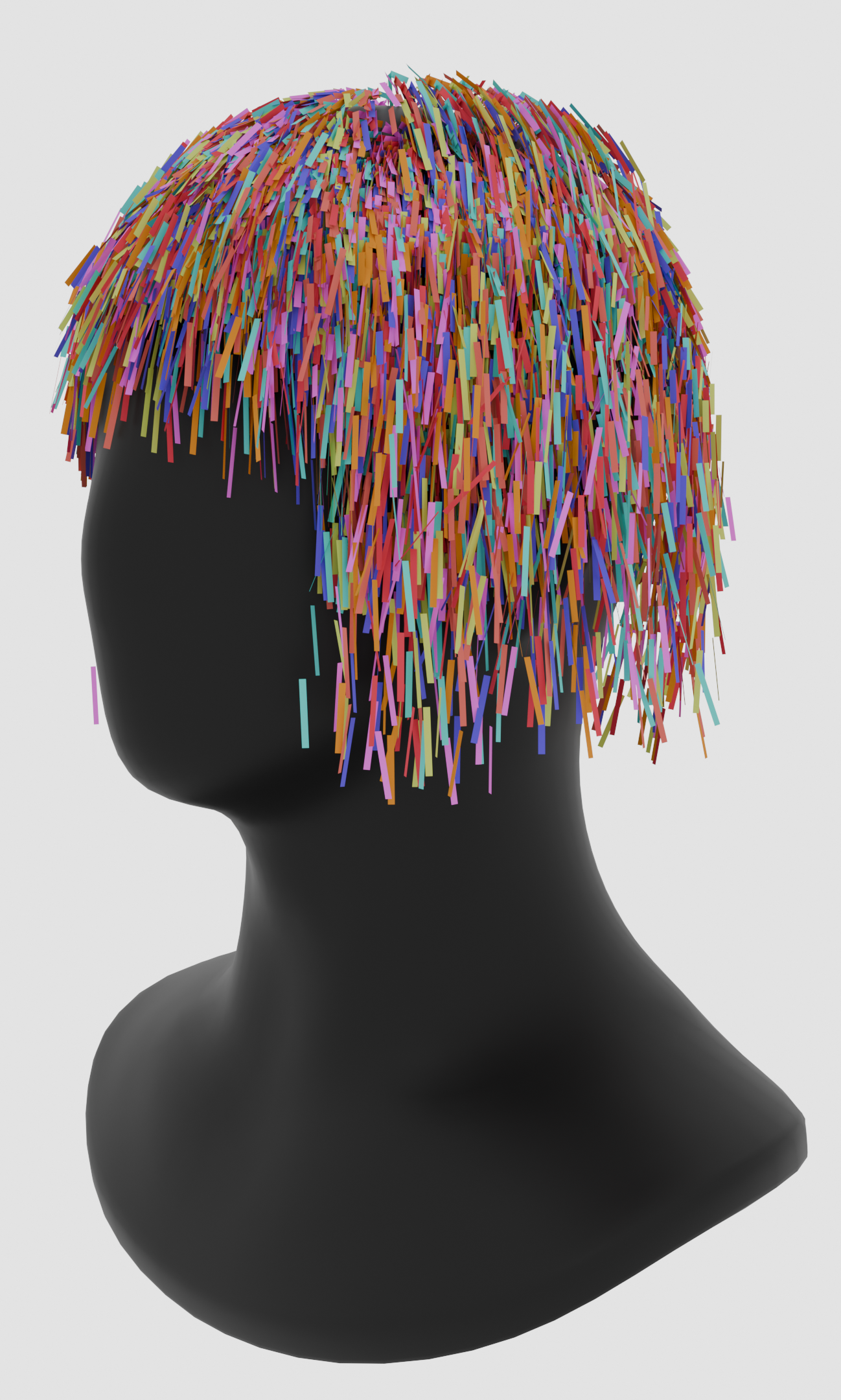}}
    \subfigure[Wavy Strands]{\includegraphics[width=0.16\linewidth]{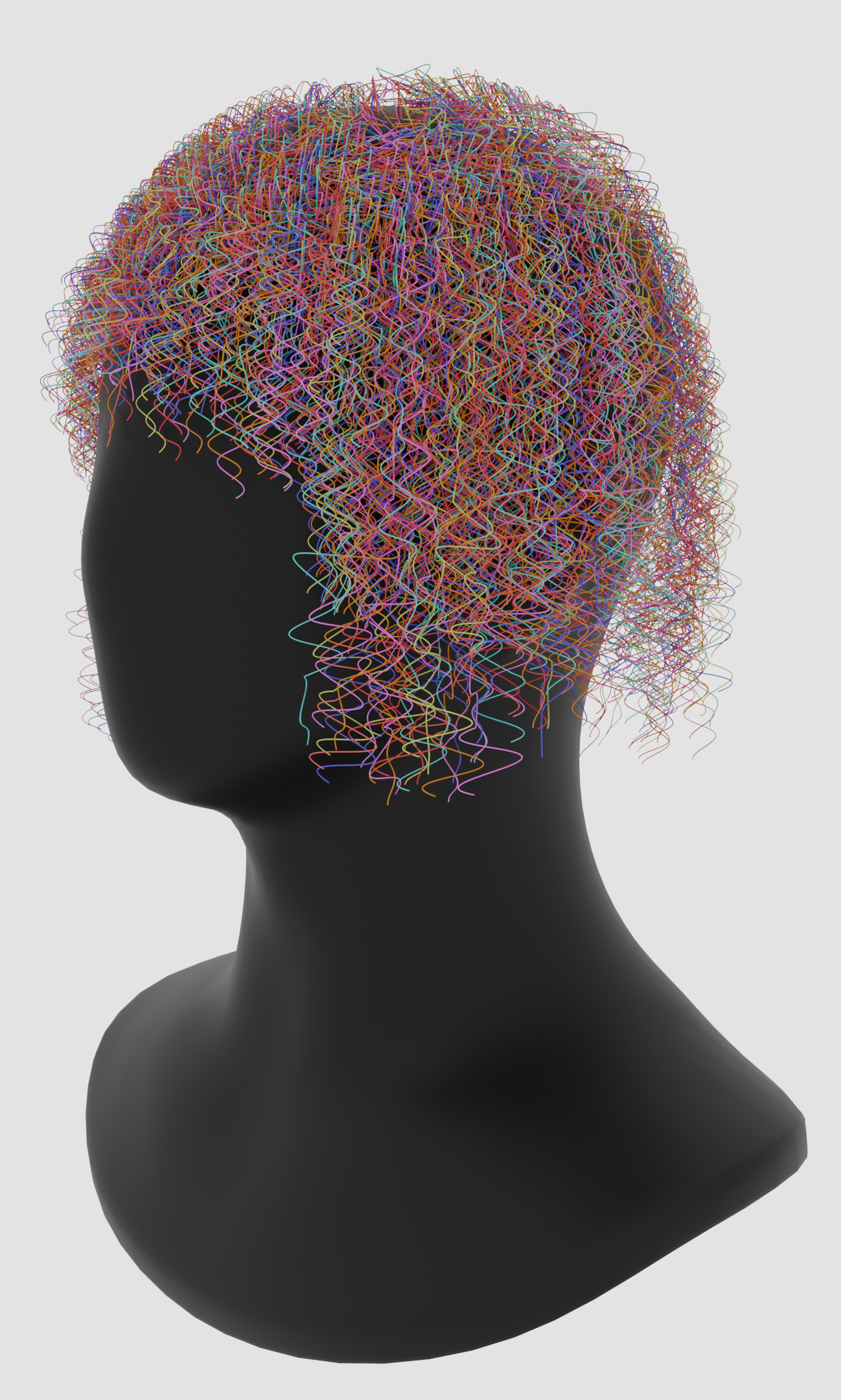}}
    \subfigure[Spiral Perm]{\includegraphics[width=0.16\linewidth]{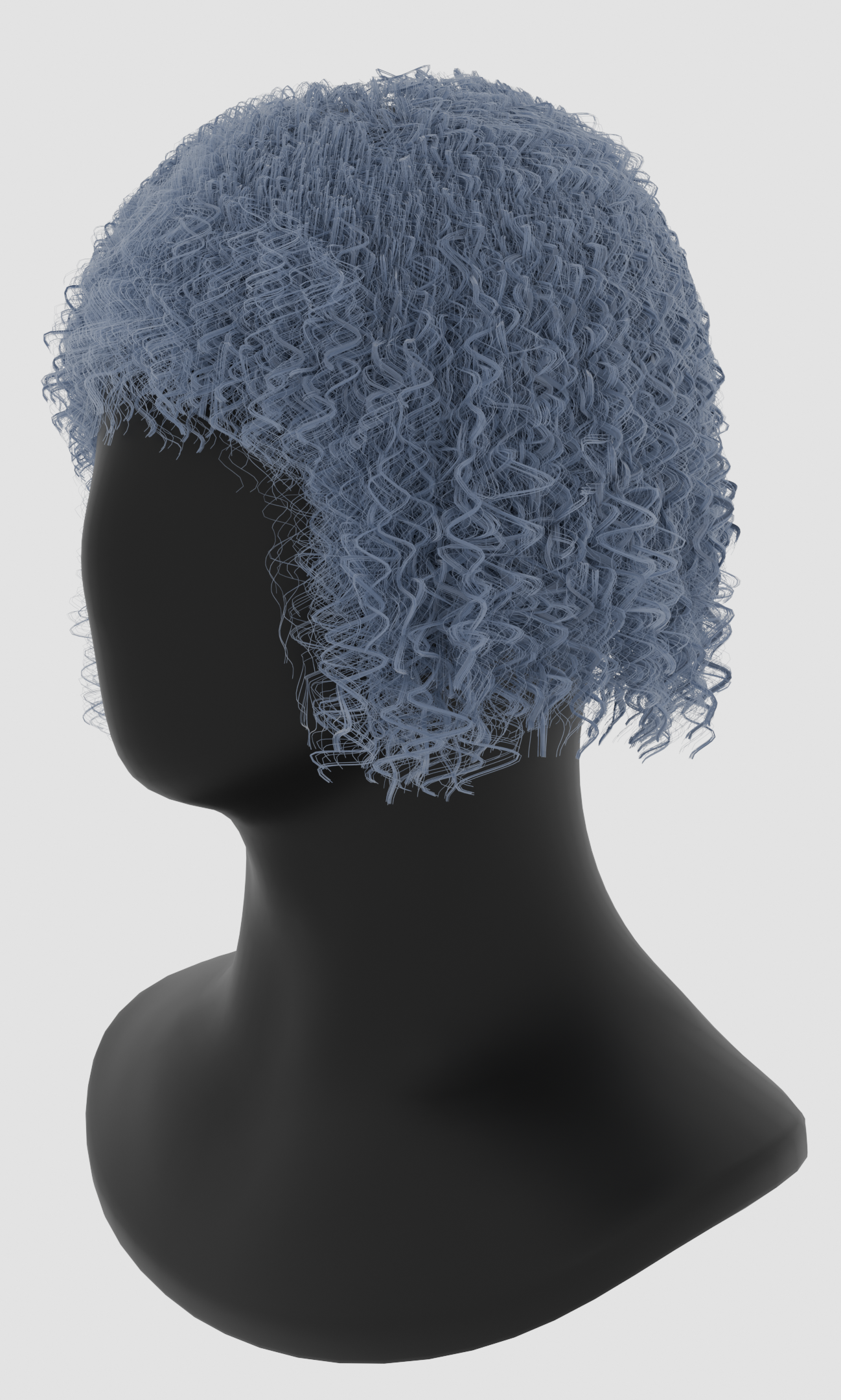}}
    % \vspace{-0.2in}
    \caption{Our curly hair model efficiently captures high-frequency features such as waves and helices while maintaining low computational cost. Each strand is decomposed into a low-frequency base configuration and a high-frequency wrinkle field, analytically represented by planar waves or spatial helices. This formulation enables accurate simulation of primary strand deformation with significantly fewer degrees of freedom than conventional rod-based methods.}
    \label{fig:teaser}
\end{teaserfigure}

%     \centering
%     \subfigure[]{\includegraphics[width=0.33\linewidth]{figures/
%     teaser1-clip.png}}
%     \subfigure[]{\includegraphics[width=0.33\linewidth]{figures/teaser2-clip.png}}
%     \subfigure[]{\includegraphics[width=0.33\linewidth]{figures/teaser3-clip.png}}
%     \caption{Our wave-based hair simulation method effectively captures high-frequency details such as curls and waves while maintaining computational efficiency.}
%     \label{fig:teaser}

\maketitle

\section{Introduction}

Strand-based hair simulation is a core component of computer graphics, underpinning realistic hair dynamics in film, games, and virtual reality. Curly geometries, including waves and helices, are critical to visual richness and realism, yet they remain particularly challenging to simulate with high fidelity. Curly hair strands exhibit complex mechanical behavior dominated by bending and twisting, with high-frequency geometric details that are difficult to capture using conventional strand-based methods. Accurately resolving these features typically requires very fine spatial discretization, incurring substantial computational cost and complicating collision handling. 

Existing methods for simulating hair with high-frequency curly details can be broadly categorized into strand-based approaches and guide-strand interpolation techniques. Strand-based methods, such as those based on discrete elastic rods (DER) \cite{Bergou2008}, provide physically grounded models that accurately capture elastic, bending, and twisting behaviors, with additional complexity arising from collisions. However, resolving waves and helices requires fine spatial discretization, leading to high computational cost and significant challenges in collision detection and response. Moreover, conventional strand-based simulations often fail to preserve high-frequency details under dynamic motion, resulting in visual artifacts and reduced realism. Guide-strand interpolation techniques \cite{ward2007survey} improve efficiency by simulating a sparse set of guide strands and generating dense follower strands through interpolation. While computationally efficient, this strategy does not enrich individual strand detail, particularly for highly irregular curl patterns with rapidly varying intrinsic curvature and torsion. In addition, many existing methods inadequately capture the coupling between macroscopic deformations (e.g., stretching) and microscopic features (e.g., curl amplitude), leading to oversmoothing and loss of characteristic spring-back behavior during straightening.

We observe that wavy and spiral hair strands exhibit inherent periodicity in their high-frequency structure. Moreover, global stretching in curly hair arises primarily from the straightening of intrinsic wrinkles, with the number of wrinkles or helical turns remaining invariant during deformation except in the fully straightened limit. This behavior indicates that bending energy governs the mechanical response of wavy hair, whereas torsional energy dominates the dynamics of spiral hair. Motivated by these observations, we propose a curly hair simulation framework that explicitly models high-frequency geometric details, such as waves and helices, while maintaining computational efficiency. Each strand is decomposed into a low-frequency base configuration and a high-frequency wrinkle field, analytically represented as planar waves or volumetric helices. This decomposition enables coarse rod-based simulation of global deformation while capturing fine-scale behavior through an efficient analytical energy model. By expressing wave and helix amplitudes as functions of macroscopic strain, the method preserves wrinkle count during deformation and yields consistent modulation of curl tightness under tension.

We further introduce a curvature-energy splitting scheme that decomposes contributions from stretching, buckling, and bending, improving physical interpretability and numerical conditioning. To enhance efficiency, we derive accurate approximations of curvature energies that significantly reduce computational cost without sacrificing visual fidelity. Collision handling is accelerated via a hybrid strategy that combines coarse collision proxies for the base configuration with analytical treatment of high-frequency details, substantially reducing collision checks. Finally, we adapt interpolation techniques to the proposed curly finite-element representation, enabling efficient synthesis of dense strands from sparse guides while preserving high-frequency structure. Together, these components enable efficient, stable, and visually faithful simulation of curly hair, as demonstrated across a range of examples.

\section{Related Work}
Existing hair simulation research encompasses particle-based models, continuum and rod-based formulations, mesh-based representations, and interpolation-driven approaches, with growing emphasis on the modeling of curly and tightly coiled hair.

\emph{Mass–Spring-Based Models.} Early hair simulation predominantly relied on mass–spring models for their simplicity and efficiency. Hair was modeled as particle chains to capture large-scale motion \cite{hadap2001modeling}, with later work improving stability via loosely connected particles \cite{bando2003animating}. \citet{Selle2008} advanced these models with robust collision handling for production use. Nevertheless, mass–spring approaches poorly represent bending and twisting, especially for highly curly hair, and typically require extensive parameter tuning.

\emph{Rod-Based and Continuum Mechanics Approaches.} To improve physical fidelity, rod-based formulations grounded in continuum mechanics have been widely adopted. Super-helices \cite{Bertails2006superhelices} introduced analytical models for intrinsically curved hair, while Cosserat rod formulations \cite{Spillmann2007} provided a principled framework for simulating one-dimensional elastic objects. Discrete elastic rods \cite{Bergou2008} further advanced this line of work by accurately modeling bending and twisting via discrete geometry, later extended to viscous effects with discrete viscous threads \cite{Bergou2010,Dominik2015}. Subsequent efforts improved numerical stability and efficiency, including position-based elastic rods \cite{Umetani2015}, ADMM-based interactive solvers \cite{Daviet2023hair}, and stable Cosserat rod formulations \cite{Hsu2025}. Nonetheless, these methods typically require fine discretization to resolve high-frequency features such as waves and helices, leading to high computational cost and challenging collision handling.

\emph{Curly and Highly Coiled Hair Models.} 
Simulation of curly and tightly coiled hair has received growing attention. \citet{Iben2013} introduced an artist-friendly framework emphasizing controllability over strict physical accuracy. More recent methods explicitly model intrinsic curvature and periodic structure, including \emph{Lifted Curls} \cite{Shi2023}, which captures curl lifting via structured geometric representations, and \emph{Curly-Cue} \cite{Wu2024}, which employs compact abstractions for efficient modeling of highly coiled strands. Nonetheless, these approaches typically rely on specialized representations and offer limited coupling between macroscopic deformation and microscopic curl response. More broadly, \citet{Darke2024} emphasized the cultural importance and technical challenges of faithfully representing Black hair, highlighting the need for physically grounded and inclusive models of diverse curl patterns.

\emph{Guide-Strand Interpolation and Neural Methods.} 
To balance efficiency and visual fidelity, many hair systems adopt guide-strand interpolation, in which a sparse set of simulated guides drives dense follower strands \cite{ward2007survey}. Extensions such as dynamically controlled interpolation \cite{Somasundaram2015} adapt follower motion to deformation, while recent data-driven approaches employ neural interpolation \cite{Lyu2022} and physically guided neural models \cite{Hsu2024,Hsu2023} for real-time performance. Despite their efficiency, these methods largely propagate existing motion rather than enriching per-strand geometry, limiting their ability to faithfully reproduce highly irregular or tightly coiled curl structures.

\emph{Real-Time and GPU-Based Hair Simulation.} For interactive applications, early GPU-based pipelines enabled real-time hair simulation and rendering via parallel processing \cite{Tariq2008,Koh2001,Volino2004}. Subsequent work improved performance through reduced representations and efficient solvers, including real-time hair meshes \cite{Wu2016,Yuksel2009} and ADMM-based formulations for interactive rod simulation \cite{Daviet2023hair}. Robust collision handling was advanced by hybrid iterative solvers for Coulomb friction \cite{Daviet2011hair,Fei2021,Han2019} and example-driven techniques that leverage simulated exemplars \cite{Hu2014robusthair}. Although real-time methods achieve substantial speedups, they typically rely on aggressive simplifications that degrade high-frequency curl fidelity, with collision handling remaining a primary computational bottleneck, particularly for finely discretized curly hair.

These limitations motivate methods that explicitly model periodic high-frequency features within efficient, physically grounded simulation frameworks.

\section{Hair Model}
\label{sec:hair-model}
A free hair strand can adopt a wide range of shapes determined by its length, curvature, and torsion. Different shapes correspond to distinct mechanical responses, motivating tailored modeling assumptions for each case. We review common geometric representations of hair strands and summarize the physical properties and energy formulations commonly used in hair simulation.

\begin{figure}
    \centering
    \subfigure[Near straight hairs]{\includegraphics[width=0.32\linewidth]{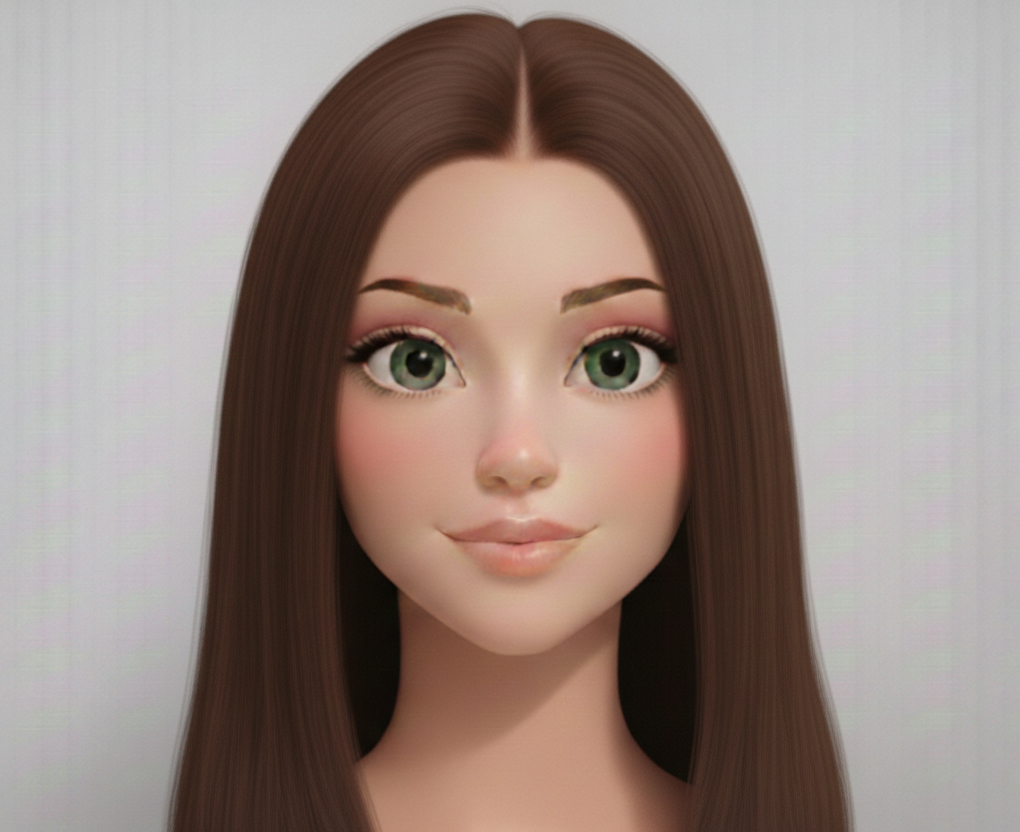}}
    \subfigure[Wavy hairs]{\includegraphics[width=0.32\linewidth]{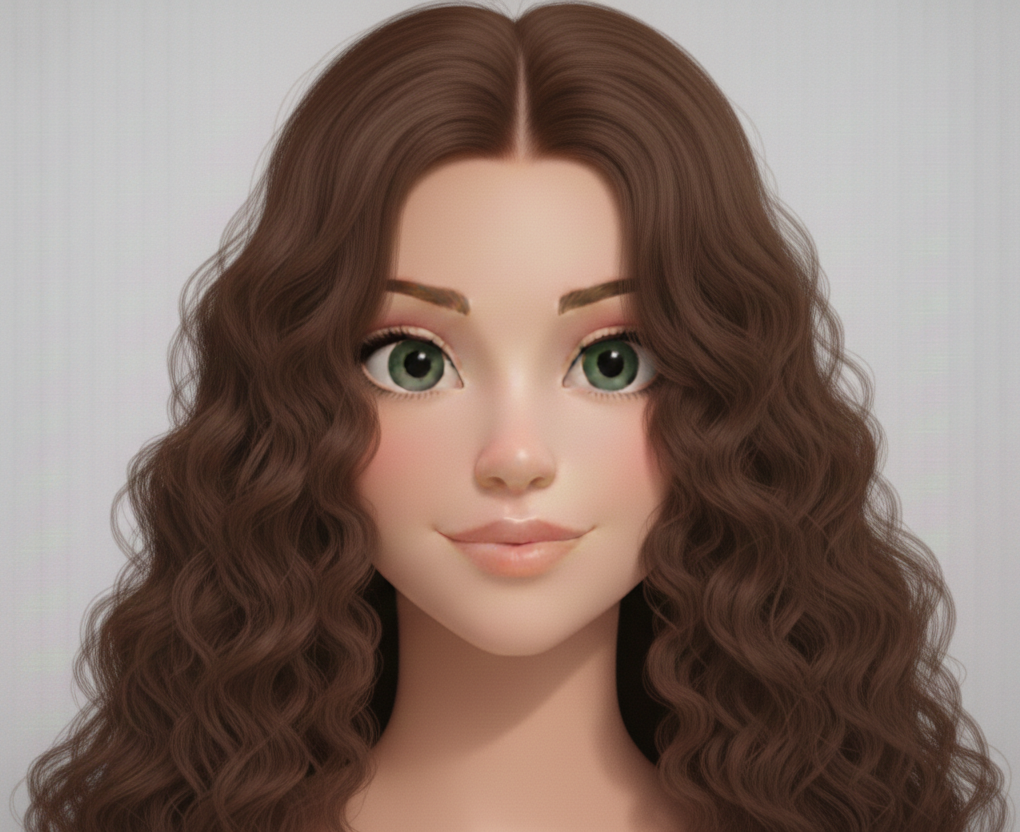}}
    \subfigure[Spiral hairs]{\includegraphics[width=0.32\linewidth]{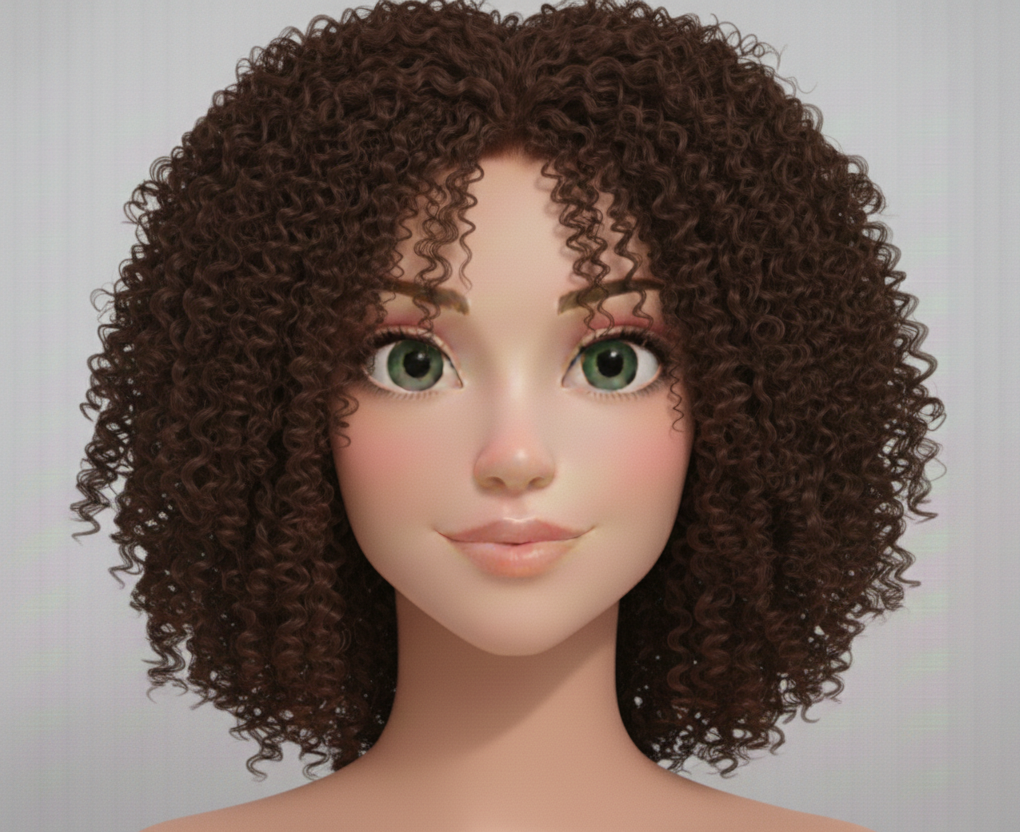}}
    \subfigure[Near straight hairs]{\includegraphics[width=0.32\linewidth]{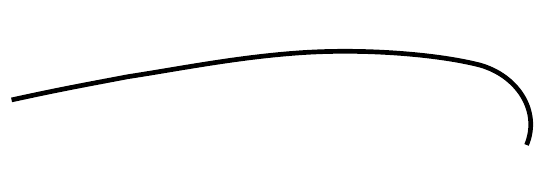}}
    \subfigure[Wavy hairs]{\includegraphics[width=0.32\linewidth]{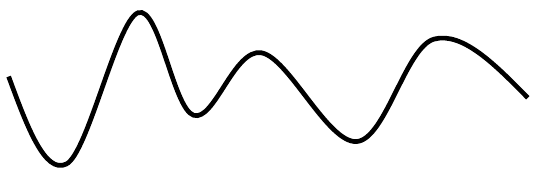}}
    \subfigure[Spiral hairs]{\includegraphics[width=0.32\linewidth]{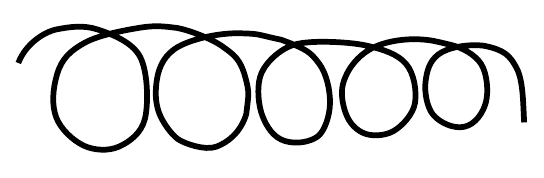}}

    \caption{Local geometry of hair strands can be classified into three categories: (a) near straight hairs, (b) wave hairs, and (c) spiral hairs.}
    \label{fig:hair-geometry}
\end{figure}

\subsection{Hair Geometry}
\label{sec:hair-geometry}
Let a hair strand be parameterized by arc length $s$ with centerline ${\bf p}(s)$. The tangent ${\bf t}(s)$, normal vector ${\bf n}(s)$, and binormal vector ${\bf b}(s)$ form a local coordinate system for geometry and deformation. Although strands may undergo complex three-dimensional deformations, local curl patterns are governed by intrinsic curvature and torsion, corresponding to natural bending and twisting, respectively. As illustrated in \autoref{fig:hair-geometry}, local hair geometry can be classified as near-straight, wavy, or spiral, dominated by stretching, bending, and torsional responses, which together determine the global shape and visual appearance of curly hair.

\subsection{Hair Physics}
\label{sec:hair-physics}

A hair strand is one-dimensional material that locally exhibits limited physical behaviors. Simply put, hair strands can be viewed as curves in three-dimensional space, and they exhibit small extensibility but obvious bending and twisting deformations. To accurately simulate hair dynamics, it is essential to understand the underlying physics governing hair strand deformation.

\subsubsection{Stretching}
The stretching energy of a hair strand can be modeled using a strain energy density function that penalizes deviations from the rest length, 
\begin{equation}
    E_\text{stretch} = \frac{1}{2} K_s \int \left(\left\|\frac{\partial\mathbf{p}}{\partial s}\right\| - 1\right)^2 ds,
    \label{eq:stretching}
\end{equation}
where $K_s$ is the material stretching stiffness.

\subsubsection{Bending}
Bending allow hair strands to curve and form complex shapes. The corresponding energy is typically modeled using the difference in curvature of the strand from its rest configuration,
\begin{equation}
    E_\text{bending} = \frac{1}{2} K_b \int \left ( \kappa - \kappa_0 \right )^2 ds,
    \label{eq:bending}
\end{equation}
where $K_b$ is the material bending stiffness, $\kappa$ is the curvature of the crinkle curve, and $\kappa_0$ is the reference curvature of the crinkle curve.

\subsubsection{Twisting}
Twisting refers to the rotational deformation of hair strands around their longitudinal axis. The twisting energy quantifies the rate of change of the strand's orientation along its length,
\begin{equation}
    E_\text{twisting} = \frac{1}{2} K_t \int \left ( \beta' - \beta'_0 \right )^2 ds,
    \label{eq:twisting}
\end{equation}
where $K_t$ is the material twisting stiffness, $\beta$ is the rotation angle of the current local frame related to the twisting-free Bishop frame of the curve, and $\beta_0$ is the corresponding rest value.

\subsection{Discrete Rod}
For computational purposes, hair strands are typically discretized into a sequence of connected elements, each defined by vertices and associated physical properties. This discrete representation enables efficient evaluation of forces and energies and facilitates numerical integration of the equations of motion governing strand dynamics.

\subsubsection{Spring-based elasticity energy}
For a discretized hair strand modeled as vertices connected by edges, stretching energy is commonly approximated using spring-based elasticity. Each edge is treated as a linear spring with rest length equal to its initial endpoint distance, and the stretching energy is evaluated independently per edge as follows:
\begin{equation}
    E_{\text{stretch}, i} = \frac{1}{2} k_s (l_i - l_{i0})^2,
\end{equation}
where $k_s$ is the spring constant, $l_i$ is the current length of edge $i$, and $l_{i0}$ is the rest length of edge $i$. The total stretching energy is then the sum of the stretching energies of all edges: $E_{\text{stretch}} = \sum_{i} E_{\text{stretch}, i}$.

\subsubsection{Angle-based bending energy}
The bending energy can be modeled using angle-based formulations for each pair of adjacent edges:
\begin{equation}
    E_{\text{bend}, i} = \frac{1}{2} k_b (\theta_i - \theta_{i0})^2,
\end{equation}
where $k_b$ is the bending stiffness constant, $\theta_i$ is the current angle between edges $i$ and $i+1$, and $\theta_{i0}$ is the rest angle between these edges. The total bending energy is the sum of the bending energies of all adjacent edge pairs: $E_{\text{bend}} = \sum_{i} E_{\text{bend}, i}$.

\subsubsection{Angle-based twisting energy}
The twisting energy can also be modeled using angle-based formulations for each segment:
\begin{equation}
    E_{\text{twist}, i} = \frac{1}{2} k_t (\beta_i - \beta_{i0})^2,
\end{equation}
where $k_t$ is the twisting stiffness constant, $\beta_i$ is the current twist angle of segment $i$, and $\beta_{i0}$ is the rest twist angle of segment $i$. The total twisting energy is the sum of the twisting energies of all segments: $E_{\text{twist}} = \sum_{i} E_{\text{twist}, i}$.

\section{Our Hair Model}
\label{sec:wavy-hair-model}

A common strategy for simulating curly hair is to prescribe non-straight rest configurations with intrinsic curvature and torsion in discrete rod models, allowing strands to naturally assume wavy or helical shapes. While effective in capturing curly geometry, accurately resolving such details typically requires very fine spatial discretization, leading to a large number of degrees of freedom, increased computational cost, and complex collision handling. 

We introduce curly elements that address these limitations by decomposing each hair strand into a low-frequency base configuration and a high-frequency wrinkle field. The formulation is motivated by the observation that the apparent stretching of wavy hair is dominated by the straightening of intrinsic high-frequency wrinkles, whose count remains invariant during deformation except in the fully straightened limit, indicating a bending-dominated mechanical response. Similarly, the extension of spiral hair primarily results from the unwinding of intrinsic helical structures, with the number of turns preserved until complete straightening, implying a torsion-dominated response.

\subsection{Parametric Representation}
\begin{figure}[t]
    \centering
    \subfigure[Twisting-free wavy element]{\includegraphics[height=0.31\linewidth]{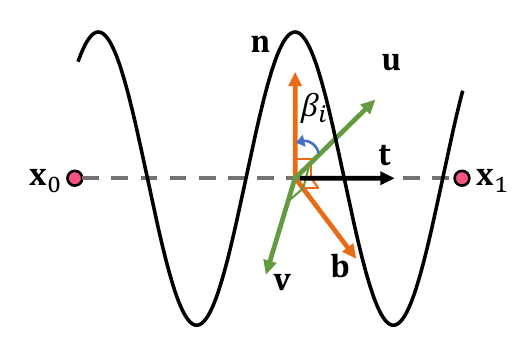}}
    \subfigure[Spiral element with twisting]{\includegraphics[height=0.31\linewidth]{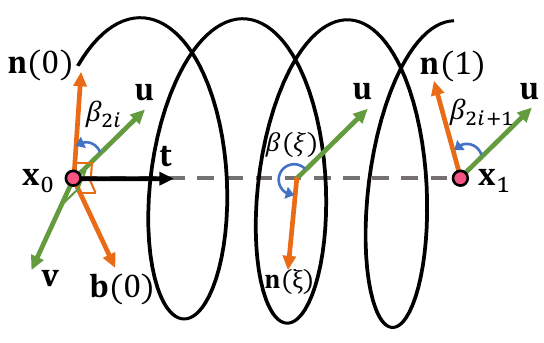}}
    \vspace{-0.1in}
    \caption{Wavy (a) and spiral (b) curves along the base curves. The base curve is represented as a polyline of connected segments, with high-frequency details modeled analytically along the normal direction of each segment.}
    \label{fig:curly_elements}
    % \vspace{-0.1in}
\end{figure}

Considering an continuous curve parameterized by arc length, i.e. ${\bf p}(s) \in \mathbb{R}^3$, where $s$ is the arc length parameter. The normal vector of the curve is denoted as ${\bf n}(s)$ which is determined by the parallel transport of the Bishop frame along the curve with or without twisting. Along this curve, we intend to represent high-frequency wavy details. Therefore, the curly curve can be represented as:
\begin{equation}
    {\bf q}_(s)={\bf p}(s)+\tau(s){\bf n}(s),
    \label{eq:wave_representation_arc}
\end{equation}
where $\tau(s)$ is the high-frequency wrinkle displacement along the normal direction. In our model, we can define the wrinkle displacement $\tau(s)$ as a constant for twisted normal to model spiral hairs or as a plane wave for untwisted normal to model wavy hairs. Specifically, for wavy hairs, we define the wrinkle displacement as:
\begin{equation}
    \tau_w(s) = A(s) \cos(\phi (s) + \phi_0),
    \label{eq: wave_function_arc}
\end{equation}
where $A(s)$ is the amplitude of the crinkles, $\phi(s)$ is the phase function, and $\phi_0$ is the phase origin. Accordingly, for spiral hairs, we define the wrinkle displacement $\tau_s(s)=A(s){\bf n}(s)$, see \autoref{fig:curly_elements}.

Because hair strands are nearly inextensible, the arc length of ${\bf q}(s)$ should be preserved during deformation. Directly enforcing this constraint on ${\bf q}(s)$ is difficult due to its high-frequency structure. Instead, we impose arc-length preservation indirectly by constraining the wrinkle displacement $\tau(s)$ and requiring the arc length of the base curve ${\bf p}(s)$ to be no greater than that of the curly curve ${\bf q}(s)$. This formulation preserves the overall strand length while retaining high-frequency wavy details.

Under this model, stretching the base curve reduces the wrinkle amplitude, producing a natural flattening effect. An explicit arc-length constraint is enforced only in the limiting case of a fully straightened strand. Moreover, this formulation reflects that axial stretching is governed primarily by bending energy for wavy hair and by torsional energy for spiral hair, both of which are dominated by the high-frequency wrinkle displacement $\tau(s)$.

\subsection{Curly Finite Elements}
\label{sec:discretization}

To enable numerical simulation, we discretize the continuous model introduced above. The base curve ${\bf p}(s)$ is represented as a polyline of $N$ segments defined by $N+1$ vertices $\{ {\bf x}_0, {\bf x}_1, \ldots, {\bf x}_N \}$, whose positions constitute the primary degrees of freedom. Each segment $e_i$ is defined by consecutive vertices ${\bf x}_i$ and ${\bf x}_{i+1}$. 

To model wavy hairs, the continuous phase field $\phi({\bf X})$, which encodes the microscopic crinkle geometry, is discretized by assigning a phase value $\phi_i$ to each vertex ${\bf x}_i$. Within each segment $e_i$, the phase is interpolated using finite-element shape functions as $\phi(\xi)=\sum N_i(\xi) \phi_i$, where $\xi \in [0,1]$ denotes the local element coordinate. The phase values are held constant throughout the simulation, ensuring preservation of the material wavenumber. Similarly, the continuous wrinkle amplitude field $A({\bf X})$ is discretized by assigning an element-wise amplitude $A_i$ to each segment $e_i$, representing the average wrinkle amplitude over that element. This variable is coupled to the macroscopic deformation and serves as a key state parameter. Consequently, the discretized wave curve is fully described by the vertex positions ${\bf x}_i$, vertex phases $\phi_i$, and element amplitudes $A_i$. Spiral hairs are modeled analogously, with the wrinkle displacement $\tau(s)$ defined as a constant helical radius per segment.

\subsection{Continuity between Discrete Elements}
The wave curve is intrinsically differentiable and exhibits $C^1$ continuity within each element. However, because adjacent elements may have distinct parameters (e.g., amplitude, wavelength, and phase), discontinuities can arise at element interfaces. To ensure that independently parameterized elements form a smooth, physically consistent curve, inter-element continuity must be enforced.

\begin{figure}[t]
    \centering
    \includegraphics[width=\linewidth]{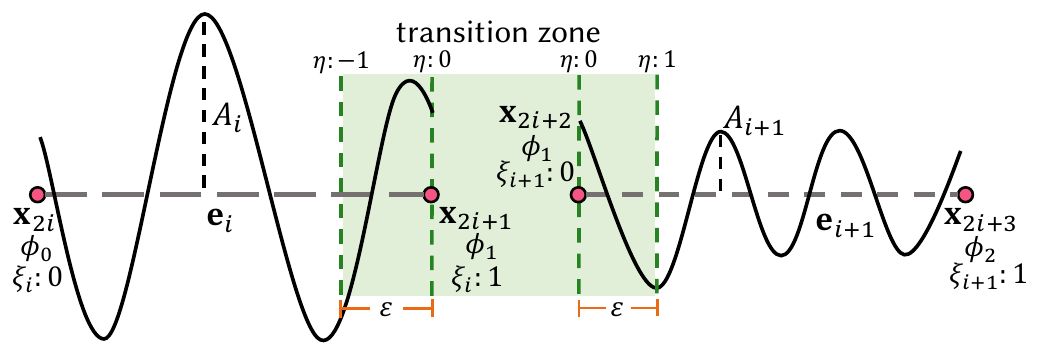}
    % \vspace{-0.1in}
    \caption{\emph{Transition zone for inter-element continuity.} A local transition zone is introduced at each element interface, where cubic Hermite interpolation ensures smooth connection of adjacent wave curves. $C^1$ continuity is initialized via amplitude and wavelength interpolation, while $C^0$ continuity and $C^1$ smoothness are maintained during simulation by matching boundary wave values and updating amplitude derivatives.}
    \label{fig:transition_zone}
    % \vspace{-0.1in}
\end{figure}

\paragraph{Initial Continuity Setup}
Discrete elements are initially connected along the base curve with independent wave parameters, leading to interface discontinuities. We address this by introducing a transition zone around each interface (\autoref{fig:transition_zone}) and enforcing smoothness via Hermite interpolation of amplitude and wavelength across neighboring elements. Phase continuity is ensured by propagating a global phase from an initial phase origin $\phi_0$ along the base curve and enforcing identical phase values at shared interfaces, yielding smooth and continuous phase evolution across elements.

\subsubsection{$C^0$ Continuity}
During simulation, wave-curve discontinuities may arise at element interfaces due to amplitude changes from stretching, axial reorientation from bending, or phase shifts induced by torsion. To preserve visual coherence and physical plausibility, we enforce at least $C^0$ continuity across element boundaries, preventing gaps or positional jumps in the wave curve.

We impose $C^0$ continuity via a nonconforming discretization of the base curve coupled with positional continuity constraints on the wave curve. For adjacent elements $e_i$ and $e_{i+1}$ sharing a logical endpoint, the corresponding wave-curve endpoints are constrained to coincide, ensuring a continuous transition (\autoref{fig:continuity}(a)). This approach decouples wave continuity from base-curve connectivity, providing increased modeling flexibility.

\subsubsection{$C^1$ Continuity} 
\begin{figure}[t]
    \centering
    \subfigure[Base-curve stretching]{\includegraphics[height=0.175\linewidth]{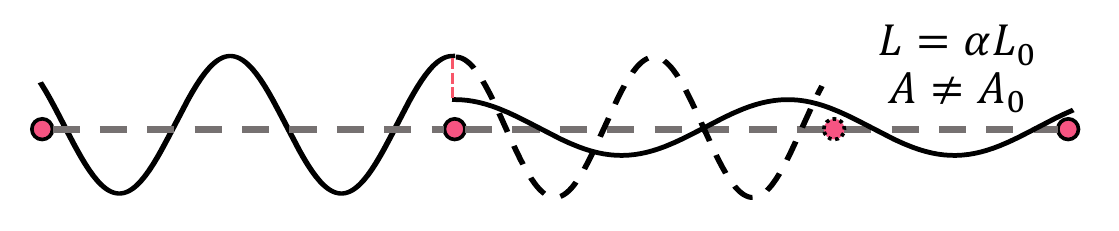}}
    \subfigure[Base-curve bending]{\includegraphics[height=0.175\linewidth]{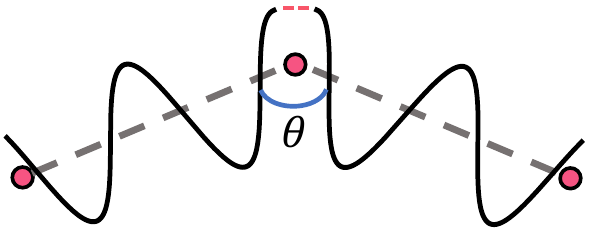}}
    \hspace{0.01\linewidth}
    \subfigure[Base-curve twisting]{\includegraphics[height=0.175\linewidth]{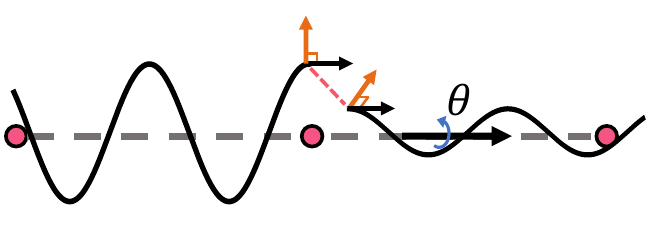}}
    % \vspace{-0.1in}
    \caption{\emph{Continuity schemes caused by base-curve deformation.} (a) During base-curve stretching, the wave amplitude decreases. During base-curve bending (b) and twisting (c), the base-curve normal direction changes, resulting gap at the element interface.}
    \label{fig:continuity}
    % \vspace{-0.1in}
\end{figure}

To enhance inter-element smoothness, we enforce $C^1$ continuity of the wave curve, which is critical for accurate bending and twisting and prevents slope discontinuities that cause visual artifacts or numerical instability. Because elements are independently parameterized and subject to a $C^0$ constraint, enforcing $C^1$ continuity is nontrivial. The wave-curve derivative depends on the base-curve position ${\bf p}$, normal ${\bf n}$, and wave amplitude $A$, all of which evolve during simulation. Since ${\bf p}$ and ${\bf n}$ are determined by base-curve deformation, we achieve $C^1$ continuity by interpolating the wave amplitude and its first derivative across adjacent elements using cubic Hermite interpolation within a predefined transition zone. At the interface of adjacent elements $e_i$ and $e_{i+1}$, let ${\bf q}'_l$ and ${\bf q}'_r$ denote the boundary tangents. Enforcing ${\bf q}'_l={\bf q}'_r$ yields the target amplitude derivatives via differentiation of \autoref{eq:wave_representation_arc}:
\begin{equation}
    A'_l {\bf n}_l \!-\! A'_r{\bf n}_r \!=\! \left( {\bf x}'_r \!-\! {\bf x}'_l \!+\! A_r {\bf n}'_r \!-\! A_l {\bf n}'_l \right)\cos^{-1}\!\phi\!-\!\left(A_r {\bf n}_r \!-\! A_l {\bf n}_l\right)\phi'\tan\phi.
\end{equation}
with derivatives taken with respect to the parametric coordinate $\xi$ at the interface. The resulting amplitude derivatives, together with boundary amplitudes, parameterize the cubic Hermite interpolation, ensuring smooth $C^1$ transitions of the wave curve (\autoref{fig:continuity}(c)).

\subsection{Curvature Energy Splitting}
\label{sec: energy_splitting}
% As noted earlier, the energy in our model is derived entirely from curvature variations of the wave curve. However, a given curvature change may result from multiple deformation modes, including base-curve stretching, bending, and buckling, making it ambiguous to attribute energy directly to curvature alone.

% To address this, we introduce an approximate splitting scheme that decomposes curvature variation into contributions from distinct deformation modes. The total curvature change is approximated as the superposition of components induced by stretching, bending, and buckling, and the total energy is expressed as the sum of corresponding mode-specific energies. This decomposition enables independent treatment of each deformation mechanism and improves the physical interpretability of the model. The total energy is thus written as
As noted earlier, the energy in our model depends solely on curvature variations of the wave curve. However, identical curvature changes may arise from different deformation modes—namely stretching, bending, and buckling—making direct curvature-based attribution ambiguous. We therefore introduce an approximate splitting scheme that decomposes the total curvature variation into mode-specific contributions. The curvature change is modeled as the superposition of stretching-, bending-, and buckling-induced components, and the total energy is expressed as the sum of the corresponding modal energies, improving physical interpretability while enabling independent treatment of each mechanism:
\[E = \frac{1}{2} B \int (\kappa - \kappa_0)^2 ds = E_{stretch} + E_{buckling} + E_{bending}.\]

\subsubsection{Stretching Energy}
\label{sec: continuous stretch energy}
\begin{figure}[t]
    \centering
    \subfigure[Reference]{\includegraphics[width=0.45\linewidth]{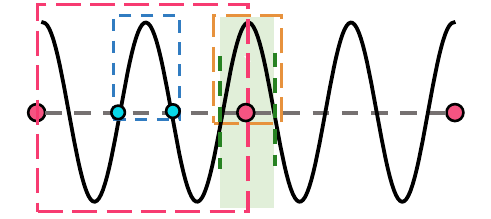}}
    \subfigure[Stretching]{\includegraphics[width=0.45\linewidth]{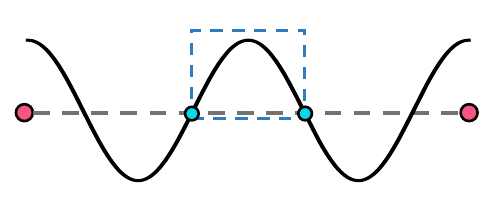}}
    \subfigure[Bending]{\includegraphics[width=0.45\linewidth]{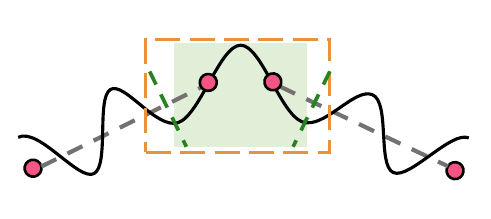}}
    \subfigure[Buckling]{\includegraphics[width=0.45\linewidth]{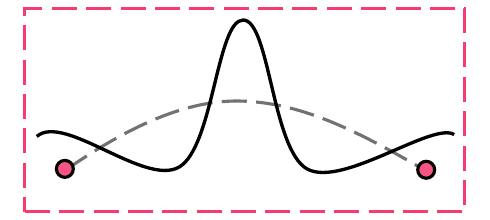}}
    \caption{The curvature energy is decomposed into mutually orthogonal components corresponding to different deformation modes. (a) Stretching energy is integrated over the entire element, approximated by summing the curvature changes of multiple half-cycle waves. (b) Buckling energy is modeled based on the change in virtual vertex. (c) Bending energy arose from local curvature variations within the transition regions between adjacent elements.}
    \label{fig:energy-splitting}
    % \vspace{-0.1in}
\end{figure}

This term models deformation induced by stretching of the base curve. In the continuous setting, it is defined by integrating curvature over the element’s wave curve. To improve efficiency, we exploit wave periodicity and approximate the stretching-induced curvature variation as the sum over multiple half-cycle waveforms (\autoref{fig:energy-splitting}(a)). The stretching energy is
\begin{equation}
    E_{stretch} =  \frac{ N_{hcw}}{2} B \int_{-\frac{\pi}{2}}^{\frac{\pi}{2}}\left( \frac{Ak^2 \cos \phi}{\left( 1 + (Ak \sin^2 \phi) \right)^{3/2}} - \kappa_0 \right)^2 ds,
\end{equation}
where $\kappa$ and $\kappa_{0}$ are the current and rest curvatures over $\phi!\in[-\pi/2,\pi/2]$, $k=1/\lambda$ is the wavenumber, $\lambda$ the wavelength, and $N_{\text{hcw}}$ the number of half-cycles per element, fixed at initialization.

Under stretching, both wavelength and amplitude evolve. We assume uniform scaling of the wavelength with base-curve length, $\lambda=(L/L_{0})\lambda_{0}$, where $L_{0}$ and $\lambda_{0}$ are the initial base length and wavelength. By arc-length conservation, the wave-curve length remains constant, requiring
\begin{equation}
    L_{wave} = \int \sqrt{1 + \left ( A k \right )^2 \sin^2(\phi)} dx=L_{wave}^0,
    \label{eq:continuous-compute-A}
\end{equation}
As this constraint has no closed-form solution and is costly to evaluate, we employ a geometric approximation (\autoref{sec:approx-stretch}) to efficiently estimate the amplitude's first derivative $A'$.

\subsubsection{Bending Energy}
Similarly, bending energy arises from curvature variations of the wave curve concentrated within the inter-element transition zone, reflecting the localized curvature induced by base-curve bending. It is defined by integrating the wave-curve curvature over this zone (\autoref{fig:energy-splitting}(c)):
\begin{equation}
    E_{bending} = \frac{1}{2} B \int_{-\varepsilon}^{\varepsilon } \left(\frac{\left \| {\bf p}'_w(\xi) \times {\bf p}''_w(\xi) \right \|}{\left \| {\bf p}''_w(\xi) \right \|^3} - \kappa_0 \right)^2 ds,
\end{equation}
where $\varepsilon$ denotes the half-width of the transition zone.

\subsubsection{Buckling Energy}
Buckling models the geometric instability of slender structures under compression, enabling storage of compressive energy through geometric reconfiguration. In our formulation, buckling allows hair strands to accommodate compression via intra-element base-curve bending without changing wrinkle frequency. Although buckling energy also originates from curvature variations of the wave curve, linear discrete elements lack the bending degrees of freedom to represent it directly. To overcome this limitation, we introduce a midpoint quadrature point per element, yielding an effective second-order representation without adding degrees of freedom, as the point is fully determined by base-curve deformation. The quadrature point is offset along the element normal to induce local curvature, converting an otherwise underconstrained higher-order formulation into a solvable one at the cost of reduced expressiveness (Fig.~\ref{fig:energy-splitting}(d)). The buckling energy is defined as
\begin{equation}
    E_{buckling} = \frac{1}{2} B \int_{-1}^{1} \left(\frac{\left \| {\bf p}'_w(\xi) \times {\bf p}''_w(\xi) \right \|}{\left \| {\bf p}''_w(\xi) \right \|^3} - \kappa_0 \right)^2 ds.
\end{equation}
where $\kappa_0$ is the rest curvature. The quadrature point position is given by ${\bf x}_{\text{qp}}=({\bf x}_{i}+{\bf x}_{i+1})/2+d{\bf n}$, with element endpoints ${\bf x}_{i},{\bf x}_{i+1}$ and normal ${\bf n}$. Enforcing arc-length conservation, the deflection $d$ is approximated geometrically as $d=\tfrac{1}{2}\sqrt{L_{0}^{2}-L^{2}}$, where $L_{0}$ and $L$ are the rest and current element lengths; $d$ vanishes under extension ($L\ge L_{0}$). This construction converts compressive deformation of a first-order element into an explicit curvature change, enabling efficient buckling energy evaluation.

\subsection{Curvature Energy Approximation}
Continuous curvature energy formulations are computationally costly and complicate Jacobian and Hessian evaluation. We therefore introduce a unified angle-based discretization that efficiently approximates the curvature energies above. Since our model is defined on the wave curve, we further apply stiffness alignment to ensure that the discrete formulation remains physically consistent with the underlying continuous curvature energy.

\begin{figure}[t]
    \centering
    \subfigure[Stretching $\theta_s$]{\includegraphics[height=0.175\linewidth]{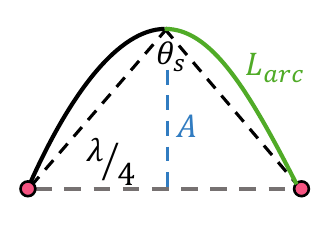}}
    \subfigure[Bending $\theta_{bd}$]{\includegraphics[height=0.175\linewidth]{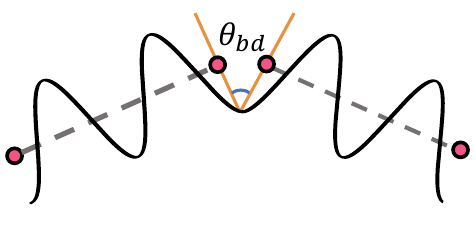}}
    \subfigure[Buckling $\theta_{bc}$]{\includegraphics[height=0.175\linewidth]{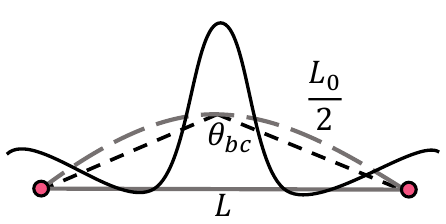}}
    \caption{A unified angle model for approximating continuous curvature energies. (a) Tangent angle over a half-cycle of the wave. (b) Deflection angle at the ghost vertex between element endpoints. (c) Normal angle at shared endpoints of adjacent elements.}
    \label{fig:theta}
\end{figure}

\subsubsection{Stretching Energy}
\label{sec:approx-stretch}
We approximate the curvature energy of each half-cycle of the wave using the change in its associated angle:
\begin{equation}
    E_{stretch} = \frac{N_{hcw}}{2} B_{s} (\theta_s - \theta_{s_0})^2,
\end{equation}
where $B_{s}$ is the stretching stiffness, $\theta_s$ is the current angle of a half-cycle wave, and $\theta_{s_0}$ is its stress-free value, see ~\autoref{fig:theta}(a).

% Evaluating $\theta_s$ requires the wave amplitude $A$ and wavelength $\lambda$, both of which vary with base-curve stretching. While $A$ can be obtained by solving the arc-length constraint in \autoref{eq:continuous-compute-A} via numerical quadrature, this approach complicates derivative evaluation. To improve efficiency, we adopt a geometric approximation in which one quarter of the waveform is modeled as a triangle. Using the Pythagorean theorem, the amplitude is approximated as 
Evaluating $\theta_s$ depends on the wave amplitude $A$ and wavelength $\lambda$, both of which vary with base-curve stretching. Solving the arc-length constraint in \autoref{eq:continuous-compute-A} via numerical quadrature yields accurate amplitudes but complicates derivative evaluation. To improve efficiency, we adopt a geometric approximation that models one quarter of the waveform as a triangle. By the Pythagorean theorem, the amplitude is approximated as
$A_\text{approx} = \sqrt{L_\text{arc}^2 - \left (\lambda/4 \right )^2}$, where $L_{\text{arc}}$ is the initial arc length of a quarter wave, computed from $\lambda_{0}$ and $A_{0}$. This approximation is used only to estimate the first derivative, yielding 
% where $L_\text{arc}$ is the initial arc length of one quarter of the wave, computed from the initial wavelength $\lambda_0$ and amplitude $A_0$. This approximation is used only to estimate the first derivative $A'$, yielding 
$A' \approx A'_\text{approx} = -\frac{\lambda}{16}A_\text{approx}^{-1}\lambda'$. 
% The angle is then given by $\theta_s = 2 \arctan \left( \frac{\lambda}{4A} \right)$. To ensure consistency between the approximate discrete energy and the exact continuous formulation, we perform stiffness matching to determine the stretching stiffness $B_s$ used in the discrete energy model:
The stretching angle is then given by $\theta_s = 2\arctan\!\left(\frac{\lambda}{4A}\right)$.
To ensure consistency between the discrete approximation and the continuous formulation, we perform stiffness matching to determine the stretching stiffness:
\begin{equation}
    B_s = \frac{1}{(\theta_s - \theta_{s0})^2 L_\text{wave}^0}\int_{-\frac{\pi}{2}}^{\frac{\pi}{2}}
        \left( \frac{Ak^2 \cos \phi}{\left( 1 + (Ak \sin^2 \phi) \right)^{3/2}} - \kappa_0 \right)^2 ds.
\end{equation}
This approximation avoids expensive integral evaluations in simulation, reducing the stretching energy to simple algebraic operations and significantly improving computational efficiency.
% This approximation eliminates complex integral evaluations, reducing the stretching energy to simple algebraic operations and substantially improving computational efficiency.

% Evaluating $\theta_s$ for each half-cycle of the wave. As mentioned in Sec. \ref{sec: continuous stretch energy}, this angle depends on the wave's amplitude $A$ and wavelength $\lambda$, which dynamically change as the base-curve elements stretch. We can compute $A$ by \autoref{eq:continuous-compute-A} by using second-order quadrature to evaluate the integral and solve a nonlinear equation for $A$. However, this approach makes the computation of the the first derivative of $A$ complicated and inefficient.

% To obtain an efficient solution, we employ a approximation: one quarter of the waveform is approximated by a triangle. Therefore, the relationship between the amplitude and the wavelength can be derived using the Pythagorean theorem:
% \begin{equation}
%     A_\text{approx} = \sqrt{L_\text{arc}^2 - \left ( \frac{\lambda}{4} \right )^2}.
% \end{equation}
% where $L_\text{arc}$ is the initial arc length of one quarter of the wave, which can be computed from the initial wavelength $\lambda_0$ and $A_0$. This approximated $A_\text{approx}$ is only used to compute the first derivative $A'\approx A'_\text{approx}=-\frac{\lambda}{16}\left(L_\text{arc}^2 - \left ( \frac{\lambda}{4} \right )^2\right)^{-\frac{1}{2}}\lambda'$. The angle is $\theta_s=2 \arctan \left( \frac{\lambda}{4A} \right)$. 

\subsubsection{Bending Energy}
% Analogous to bending formulations in classical finite element methods, the discrete bending energy in our model arises from the relative rotation between adjacent elements. We define the bending angle $\theta_{bd}$ as the angle between the normal vectors of the microscopic wave at the shared endpoints of neighboring elements , which approximates the curvature variation over the inter-element transition zone. The bending energy is given by
Analogous to classical finite element bending formulations, the discrete bending energy arises from the relative rotation of adjacent elements. We define the bending angle $\theta_{bd}$ as the angle between the wave normals at the shared endpoint of neighboring elements, which approximates the curvature variation across the inter-element transition zone. The bending energy is
\begin{equation}
    E_{bending} = \frac{1}{2} B_{bd} (\theta_{bd} - \theta_{bd_0})^2,
\end{equation}
where $B_{bd}$ is the bending stiffness and $\theta_{bd_0}$ the rest angle, see \autoref{fig:theta}(b). 
% The stiffness $B_{bd}$ is calibrated to match the corresponding continuous curvature energy, yielding
To ensure consistency with the continuous curvature energy, $B_{bd}$ is obtained by stiffness matching:
\begin{equation}
B_{bd} = \frac{1}{(\theta_{bd} - \theta_{bd_0})^2 L_\text{wave}^{tz}}\int_{-\varepsilon}^{\varepsilon} \left(\frac{\left \| {\bf p}'_w(\xi) \times {\bf p}''_w(\xi) \right \|}{\left \| {\bf p}''_w(\xi) \right \|^3} - \kappa_0 \right)^2 ds,
\end{equation}
where $L_\text{wave}^{tz}$ is the wave-curve length within the transition zone.

\subsubsection{Buckling Energy}
Under compression, curvature variation within an element is dominated by its overall deflection. Accordingly, the virtual angle used to model buckling energy is derived directly from this deflection. The buckling energy is defined as
\begin{equation}
    E_{buckling} = \frac{1}{2} B_{bc} (\theta_{bc} - \theta_{bc_0})^2,
\end{equation}
where $B_{bc}$ is the buckling stiffness, $\theta_{bc}$ is the current element angle, and $\theta_{bc_0}$ is its rest value, see \autoref{fig:theta}(c). The angle $\theta_{bc}$ is computed from the element endpoints as
\begin{equation}
    \theta_{bc} = \mathrm{atan2} \left( \frac{2L}{L_0} \sqrt{1 - \left( \frac{L}{L_0} \right)^2}, 1 - 2\left( \frac{L}{L_0} \right)^2 \right).
\end{equation}
In the stretched regime, both the deflection $d$ and $\theta_{bc}$ vanish. The stiffness $B_{bc}$ is calibrated to match the continuous curvature energy,
\begin{equation}
    B_{bc} = \frac{1}{(\theta_{bc} - \theta_{bc_0})^2 L_{0}}\int_{-1}^{1} \left(\frac{\left \| {\bf p}'_w(\xi) \times {\bf p}''_w(\xi) \right \|}{\left \| {\bf p}''_w(\xi) \right \|^3} - \kappa_0 \right)^2 ds.
\end{equation}

\begin{figure}[t]
    \centering
    \subfigure[Element AABB]{\includegraphics[width=0.32\linewidth]{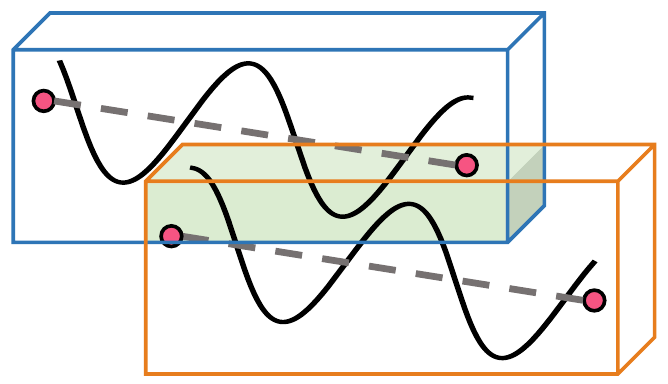}}
    \subfigure[Element Cylinder]{\includegraphics[width=0.32\linewidth]{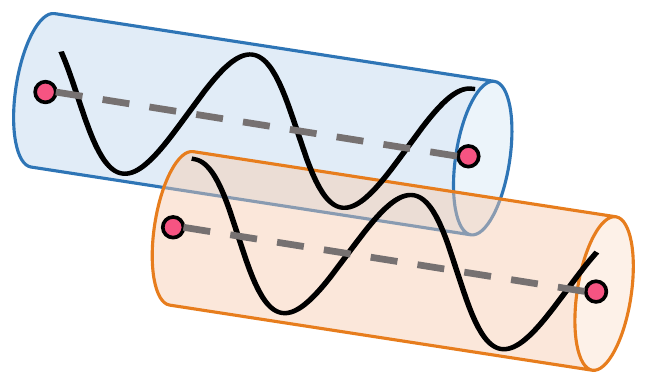}}
    \subfigure[Quarter-wave AABB]{\includegraphics[width=0.32\linewidth]{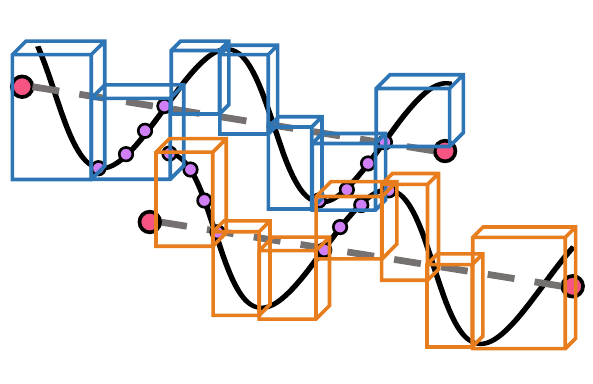}}
    \caption{Broad-phase simulation results of wavy and spiral hairs using our wave-based model. The model efficiently captures complex hair dynamics while maintaining high-frequency details.}
    \label{fig:broad_phase}
\end{figure}

\subsection{Collision Handling and Interpolation}
To ensure physical plausibility and visual realism, we integrate collision handling into our curly-element hair model using a two-phase strategy. In the broad phase, we apply multi-level culling to efficiently identify potential collisions among hair strands and between strands and external objects. Specifically, each curly element is enclosed by an axis-aligned bounding box (AABB) for coarse rejection (\autoref{fig:broad_phase}(a)), followed by a tighter bounding cylinder (\autoref{fig:broad_phase}(a)) that approximates the wave curve geometry. To further account for high-frequency structure, we additionally construct AABBs around quarter-wave segments (\autoref{fig:broad_phase}(a)). These hierarchical bounding volumes extend naturally to spiral elements and enable efficient broad-phase detection while preserving geometric fidelity. In the narrow phase, collisions are evaluated using uniformly sampled points along the wave curve parameterized by arc length, ensuring accurate detection of contacts involving high-frequency wavy and helical details. Detected collisions are resolved using impulse-based responses applied to colliding point pairs, yielding stable separation while preserving the intrinsic wave characteristics of the strands.

\begin{figure}[t]
    \centering
    \includegraphics[width=0.6\linewidth]{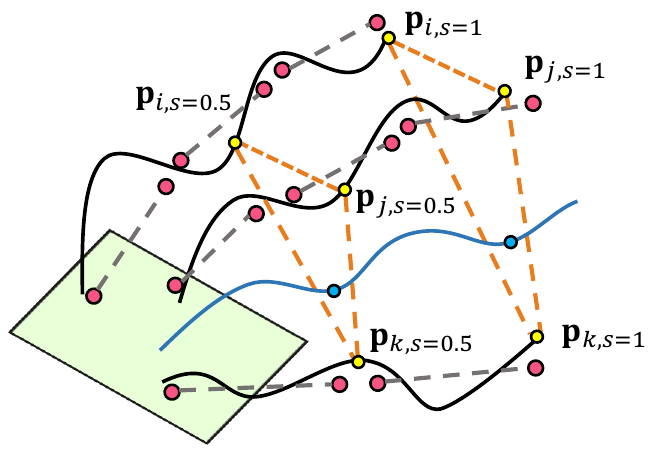}
    \caption{Interpolation of hair strands using the proposed wave-based model, enabling smooth transitions between distinct configurations while preserving high-frequency geometric details.}
    \label{fig:strand_interpolation}
\end{figure}

To improve efficiency and enrich geometric detail, we introduce an interpolation scheme built on our curly-element hair model. Given multiple guide strands represented in the wave-based formulation, intermediate strands are generated by interpolating vertex positions. For each target strand, we precompute its associated guide strands and corresponding barycentric interpolation weights, which remain fixed throughout simulation. To accommodate differing strand lengths, guide strands are first reparameterized by normalized arc length in $[0,1]$. Each vertex of the target strand is then mapped to corresponding locations on the guide strands based on this normalized parameter, from which barycentric weights are computed. During simulation, the target strand's geometry are updated by interpolating those of the guide strands using the precomputed weights. As shown in \autoref{fig:strand_interpolation}, this scheme yields smooth transitions in both macroscopic shape and microscopic wave structure, enabling efficient synthesis of dense curly hair while preserving high-frequency detail and offering flexible artistic control.

\section{Results}

In this section, we present a series of results demonstrating the capabilities of our method in capturing high-frequency wavy and spiral hair dynamics efficiently.

\begin{figure}[t]
    \subfigure[Wavy stretching]{\includegraphics[width=0.45\linewidth]{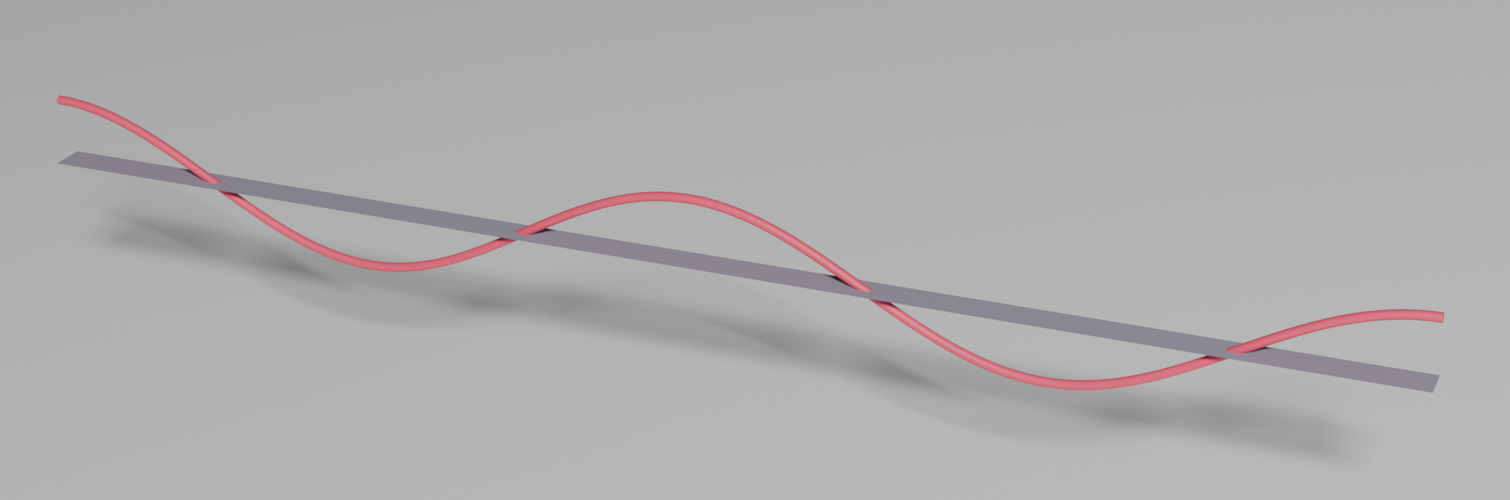}}
    \subfigure[Wavy bending]{\includegraphics[width=0.45\linewidth]{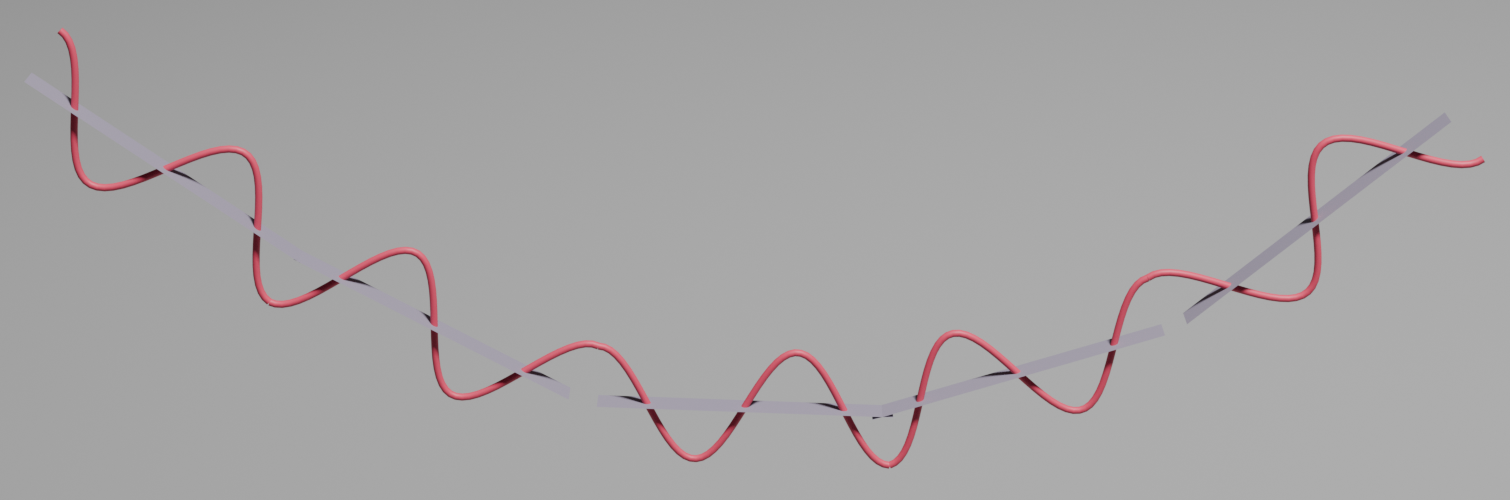}}
    \subfigure[Wavy buckling]{\includegraphics[width=0.45\linewidth]{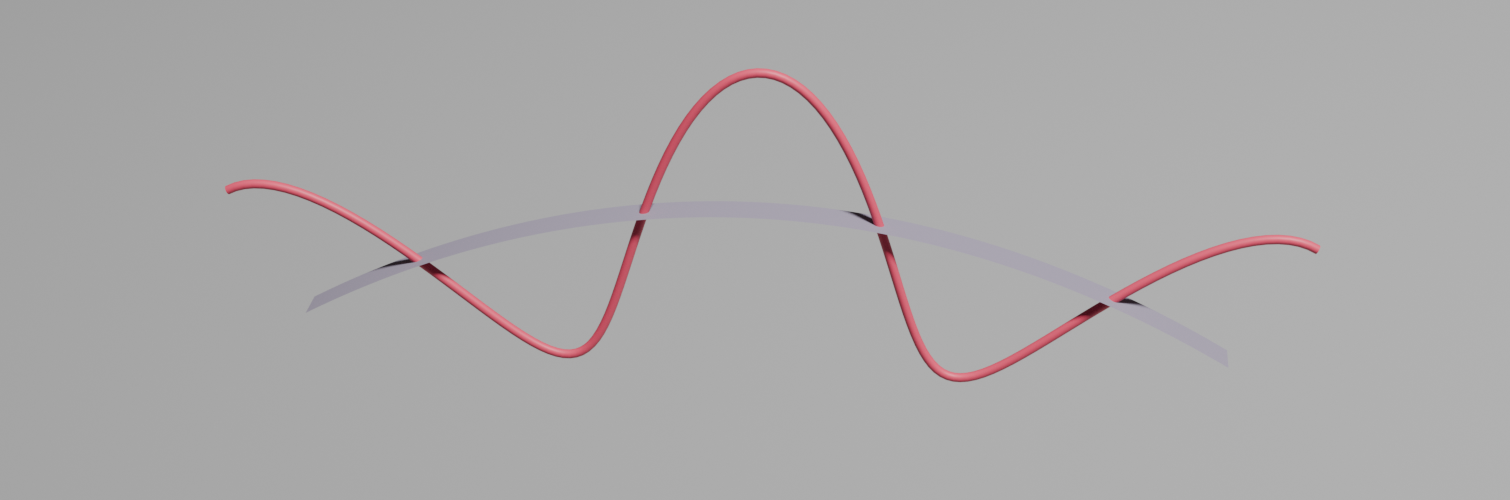}}
    \subfigure[Wavy twisting]{\includegraphics[width=0.45\linewidth]{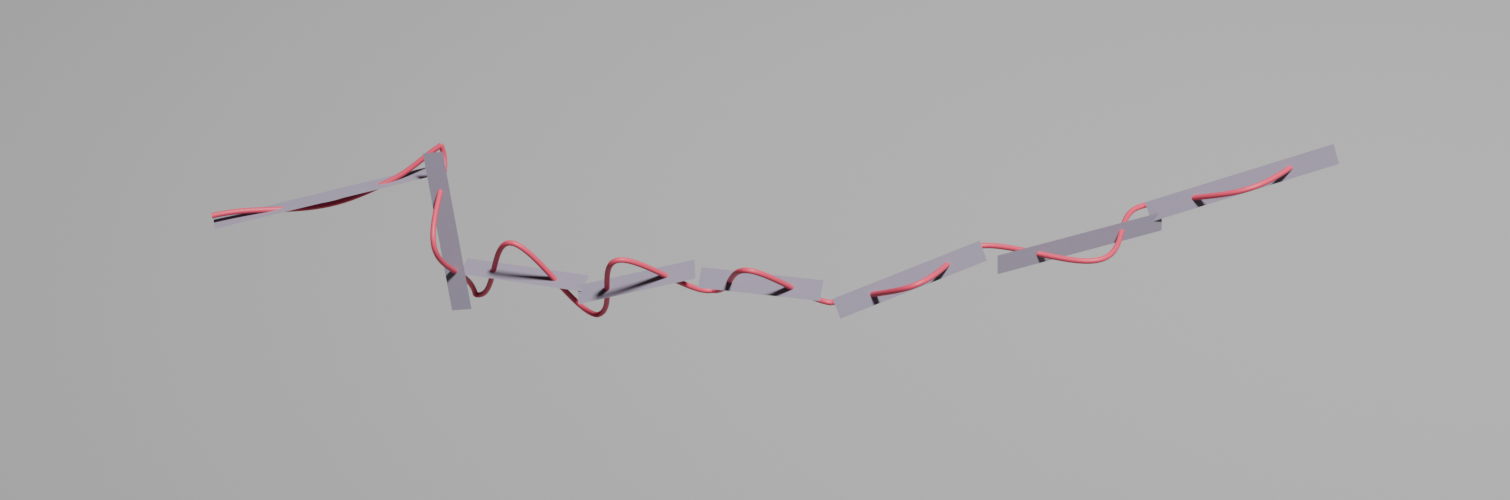}}
    \subfigure[Spiral stretching]{\includegraphics[width=0.45\linewidth]{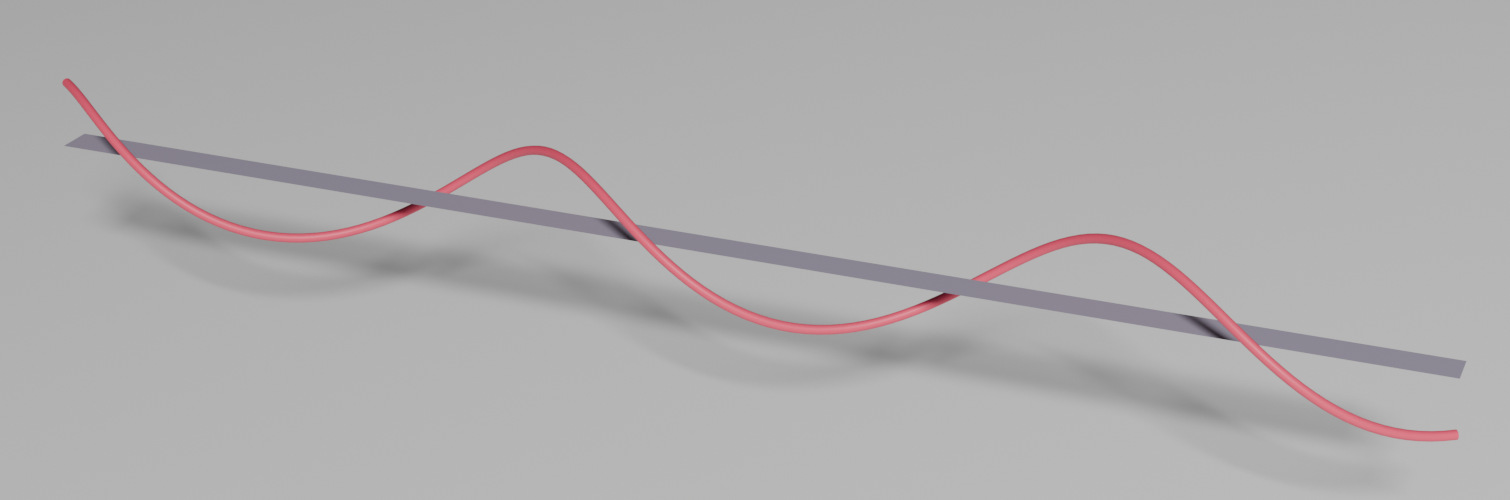}}
    \subfigure[Spiral bending]{\includegraphics[width=0.45\linewidth]{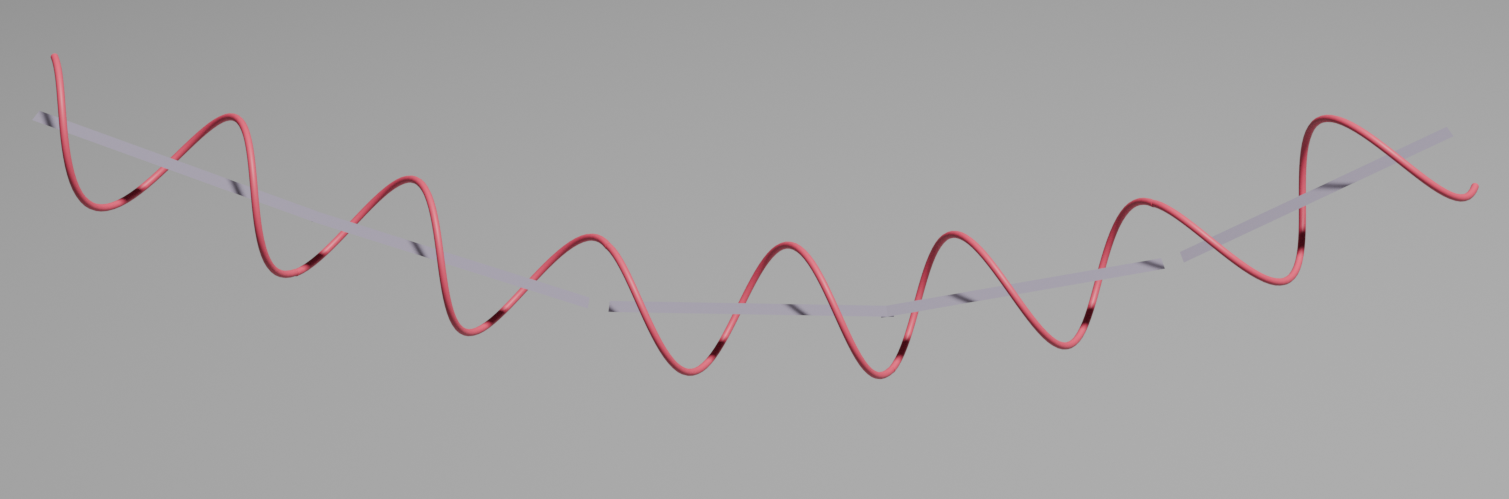}}
    \subfigure[Spiral buckling]{\includegraphics[width=0.45\linewidth]{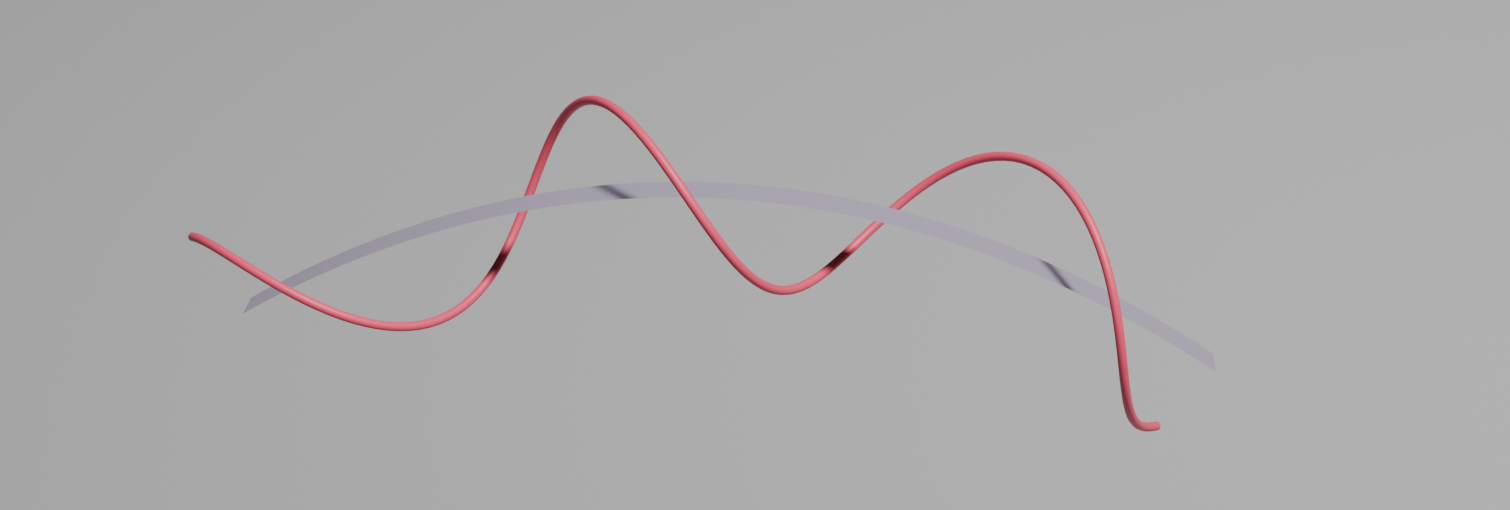}}
    \subfigure[Spiral twisting]{\includegraphics[width=0.45\linewidth]{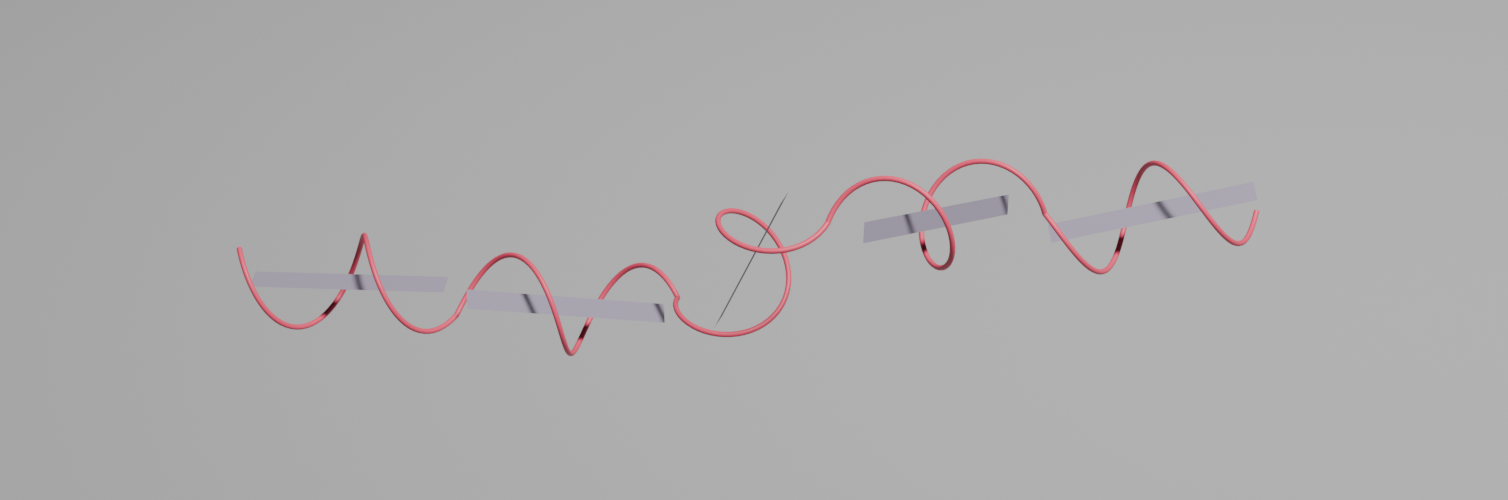}}
    \caption{Unit tests for wavy and spiral elements assess key curvature-dominated deformation modes—stretching, bending, buckling—and include torsional (twisting) behavior, validating the model’s physical accuracy and stability.}
    \label{fig:unit-test}
\end{figure}

\subsection{Unit Tests}

As shown in \autoref{fig:unit-test}, we conduct a series of unit tests to validate the accuracy and robustness of our wavy and spiral elements. These tests focus on key aspects such as energy conservation, deformation behavior under various loading conditions, and the effectiveness of the curvature energy splitting scheme. We simulate individual curly elements subjected to controlled stretching, bending, and buckling deformations. The unit tests confirm that our model accurately captures the expected physical responses, maintaining stability and fidelity across a range of scenarios. The results from these unit tests provide confidence in the reliability of our approach for simulating complex curly hair dynamics.

\begin{table*}
    \caption{We report performance statistics for various curly hair styles simulated using our curly element-based hair model and compare them with the baseline Discrete Elastic Rod (DER) method \cite{Bergou2008}. The table summarizes the number of strands (\#.S), elements (\#.E), vertices (\#.V), degrees of freedom (DoFs), and a detailed timing breakdown, including collision detection (CD), Hessian assembly (HA), the preconditioned conjugate gradient solver (PCG), and the total per-frame simulation time (TPF) in milliseconds.}
    \begin{tabular}{c | c c c c | c c c c | c c }
        \hline
        \multirow{2}{*}{Examples} & \multirow{2}{*}{\#.S} & \multirow{2}{*}{\#.E} & \multirow{2}{*}{\#.V} & \multirow{2}{*}{DoFs} & \multicolumn{4}{c|}{Performance} & \multicolumn{2}{c}{DER Comparison} \\
        \cline{6-11}
            & & & & & CD & HA & PCG & TPF & DoFs-1/TPF & DoFs-2/TPF \\ \hline
        Wavy Hair    & 546   & 6,412  & 12,824 & 45K & 7.5\% & 37.3\% & 46\% & 643.39 & 64K/756.64 & 255K/5,454  \\
        Twisted Perm & 1,143 & 10,287 & 20,574 & 72K & 11.4\% & 36\% & 42.9\% & 1,134.85 & 90K/1,249 & 230K/3,723 \\
        Spiral Perm  & 1,120 & 9,371  & 18,742 & 66K & 11.7\% & 40.3\% & 42.1\% & 1,283.97 \\
        Afro         & 1,430 & 8,313  & 16,626 & 58K & 6.4\% &  37.1\% &  43.2\% & 606.91 \\
        Spiral Hair  & 1,362 & 5,448  & 10,896 & 38K & 5.9\% & 40.5\% & 39.6\% & 335.89 \\
        \hline
    \end{tabular}
    \label{tab:performance}
\end{table*}

\subsection{Curly Hair Styles}

\begin{figure}[t]
    \centering
    \subfigure[Base Strands]{\includegraphics[width=0.32\linewidth]{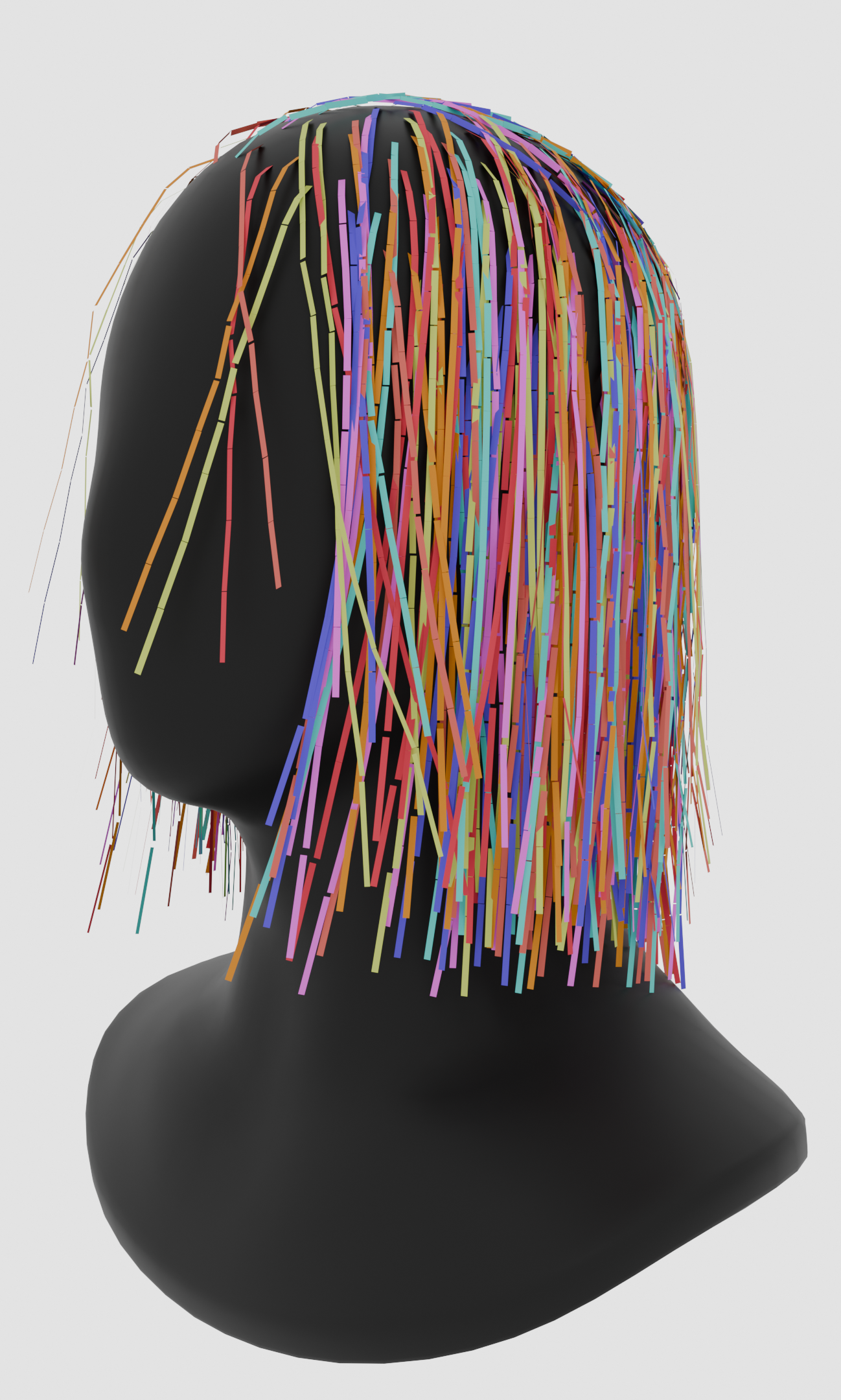}}
    \subfigure[Wavy Strands]{\includegraphics[width=0.32\linewidth]{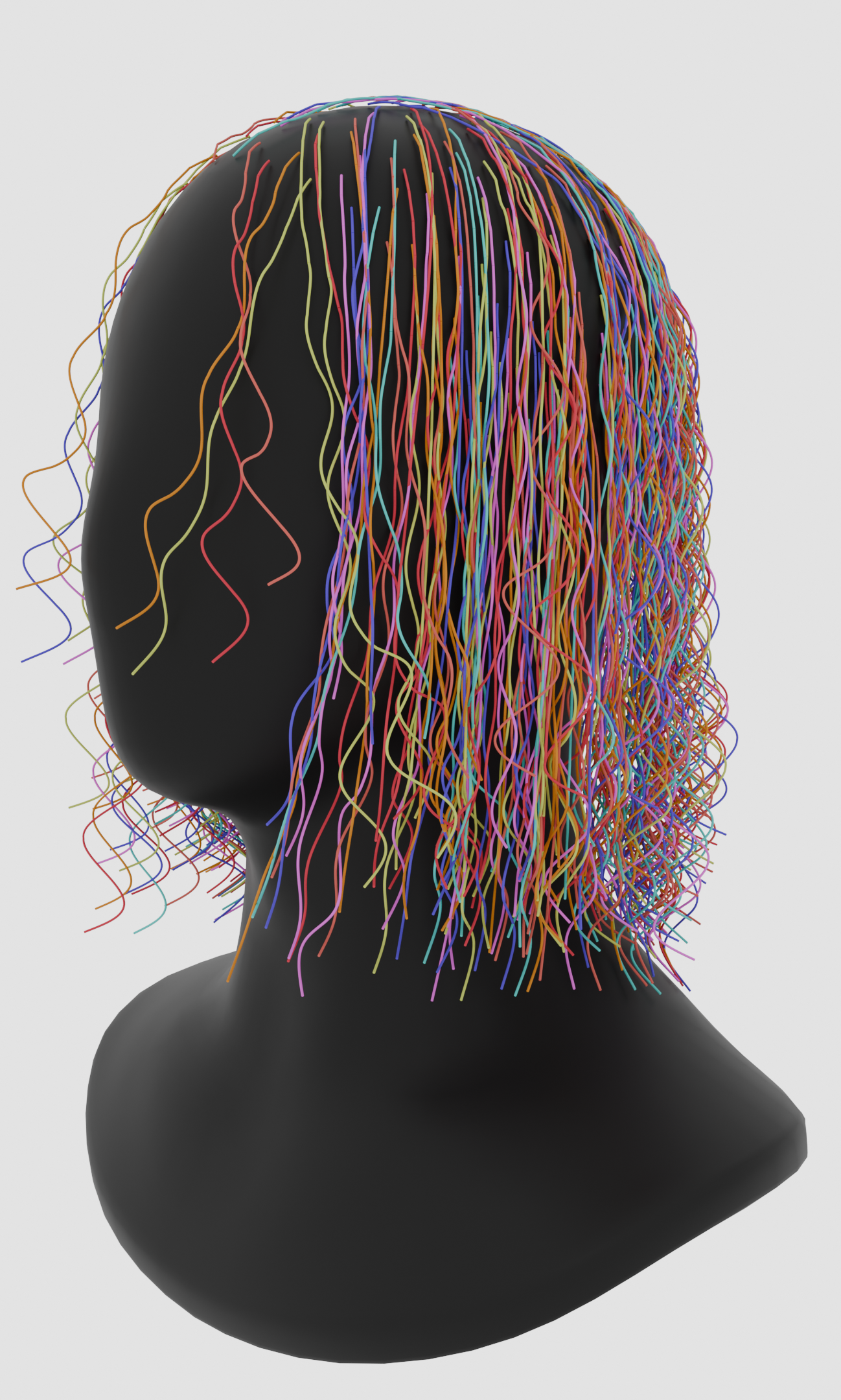}}
    \subfigure[Wavy Hair]{\includegraphics[width=0.32\linewidth]{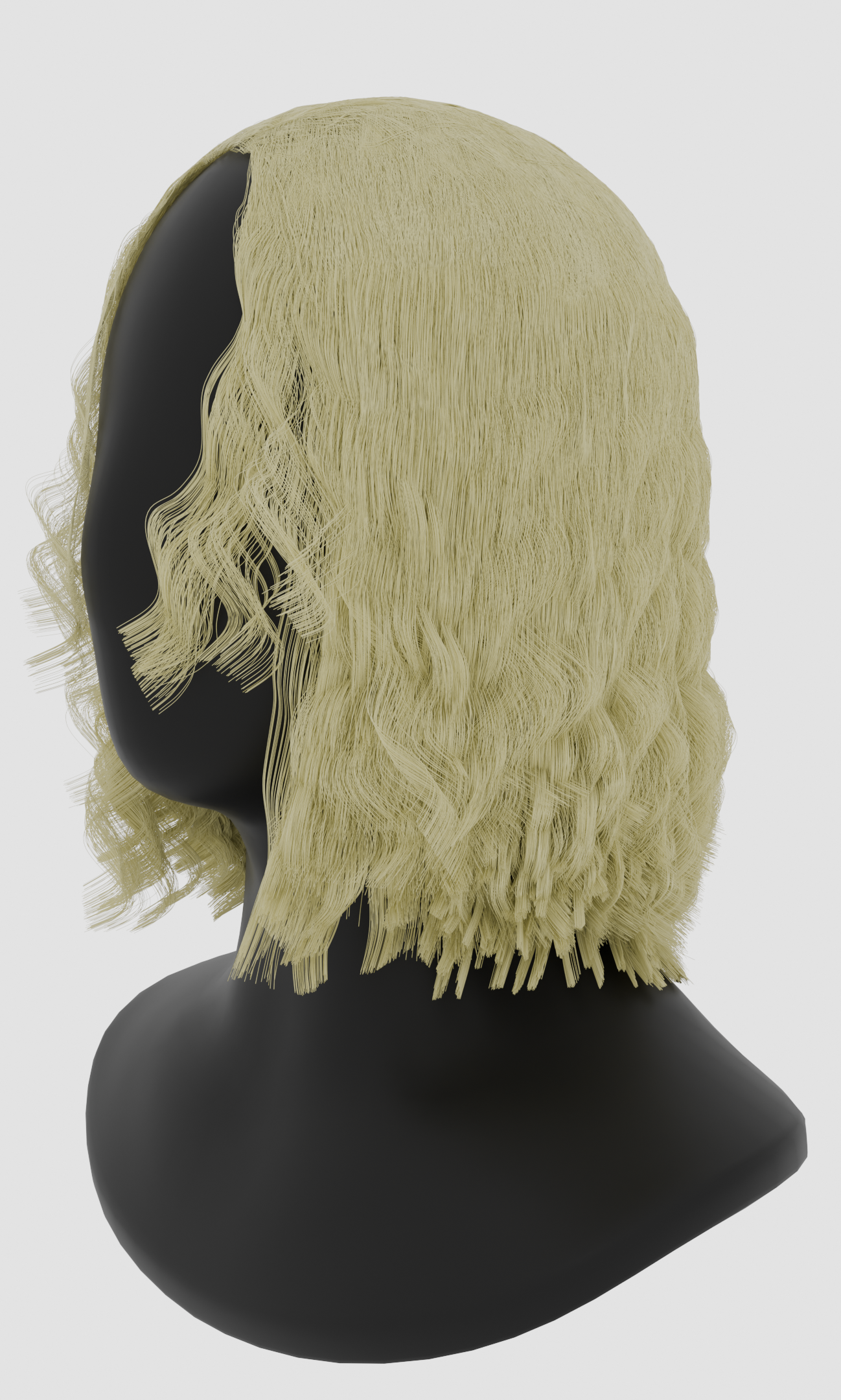}}
    \subfigure[Base Strands]{\includegraphics[width=0.32\linewidth]{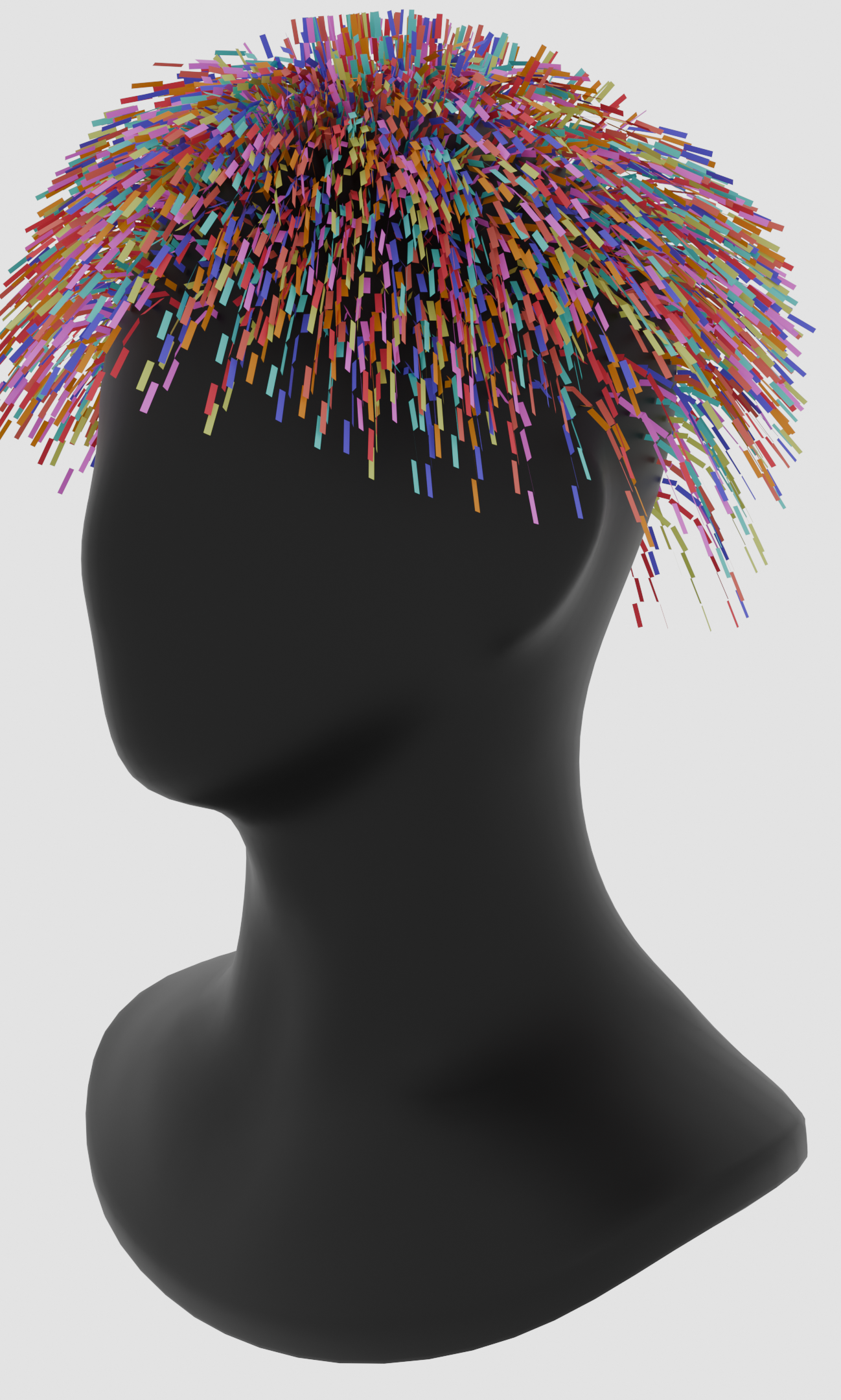}}
    \subfigure[Wavy Strands]{\includegraphics[width=0.32\linewidth]{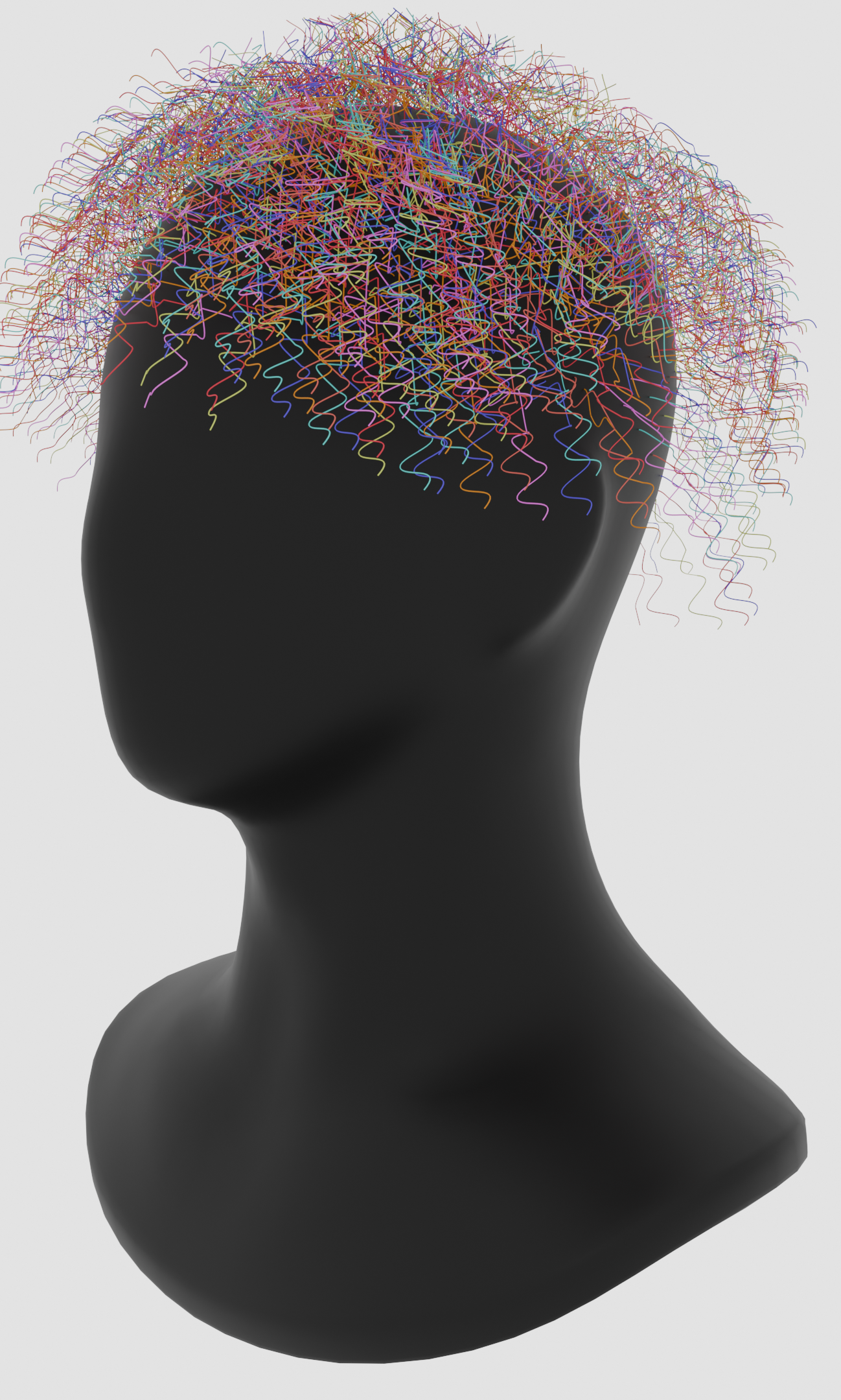}}
    \subfigure[Afro]{\includegraphics[width=0.32\linewidth]{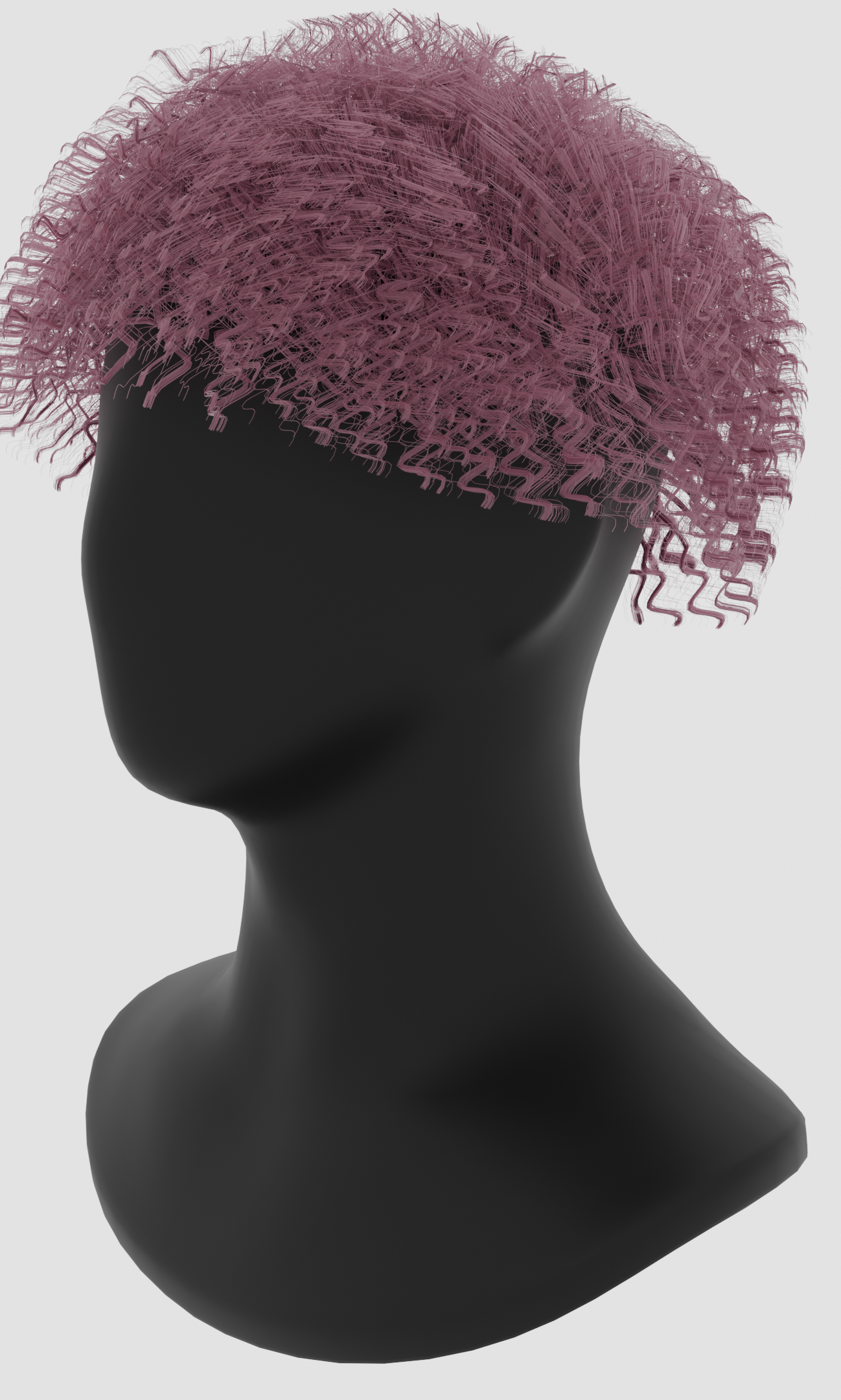}}
    \subfigure[Base Strands]{\includegraphics[width=0.32\linewidth]{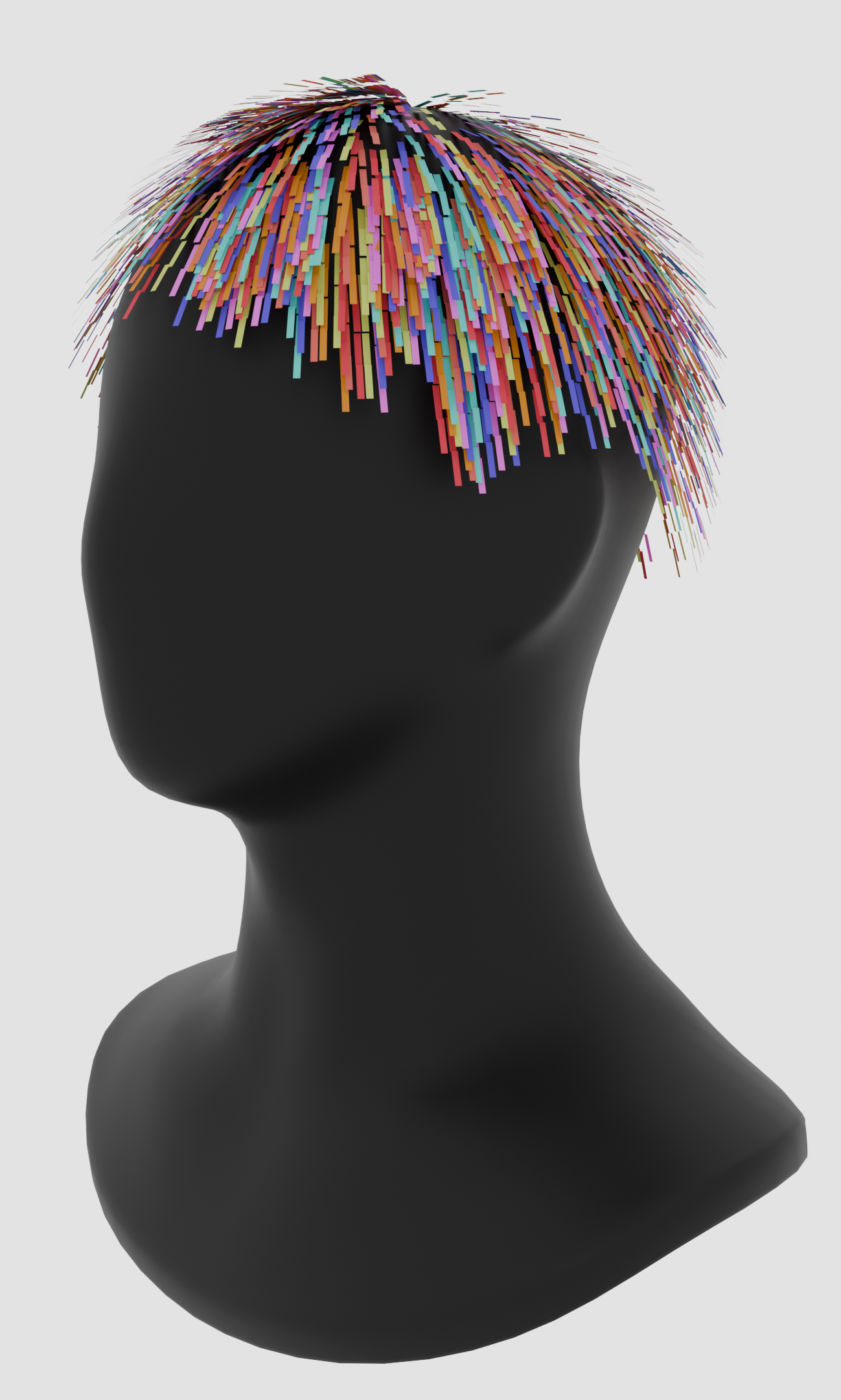}}
    \subfigure[Spiral Strands]{\includegraphics[width=0.32\linewidth]{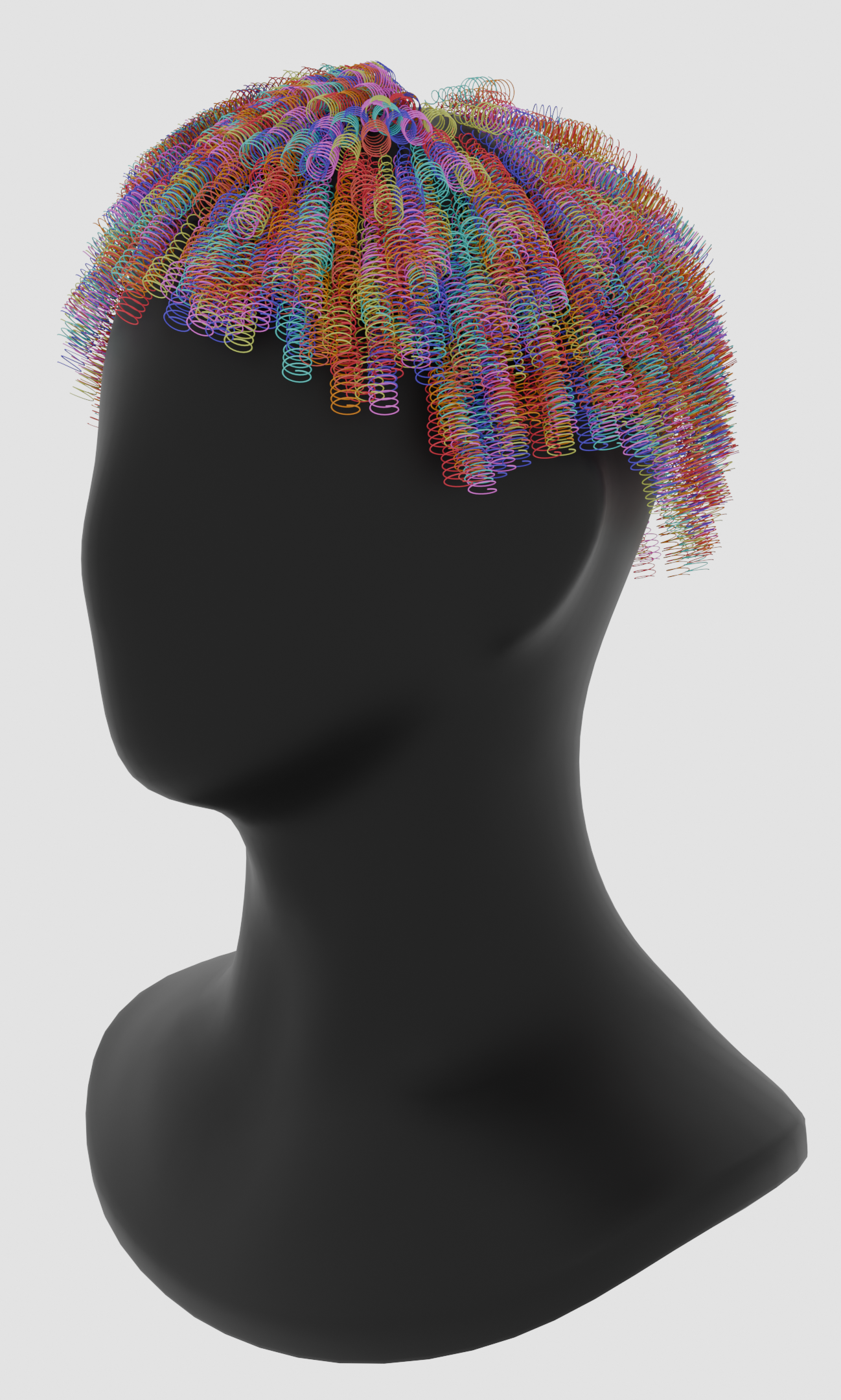}}
    \subfigure[Spiral Hair]{\includegraphics[width=0.32\linewidth]{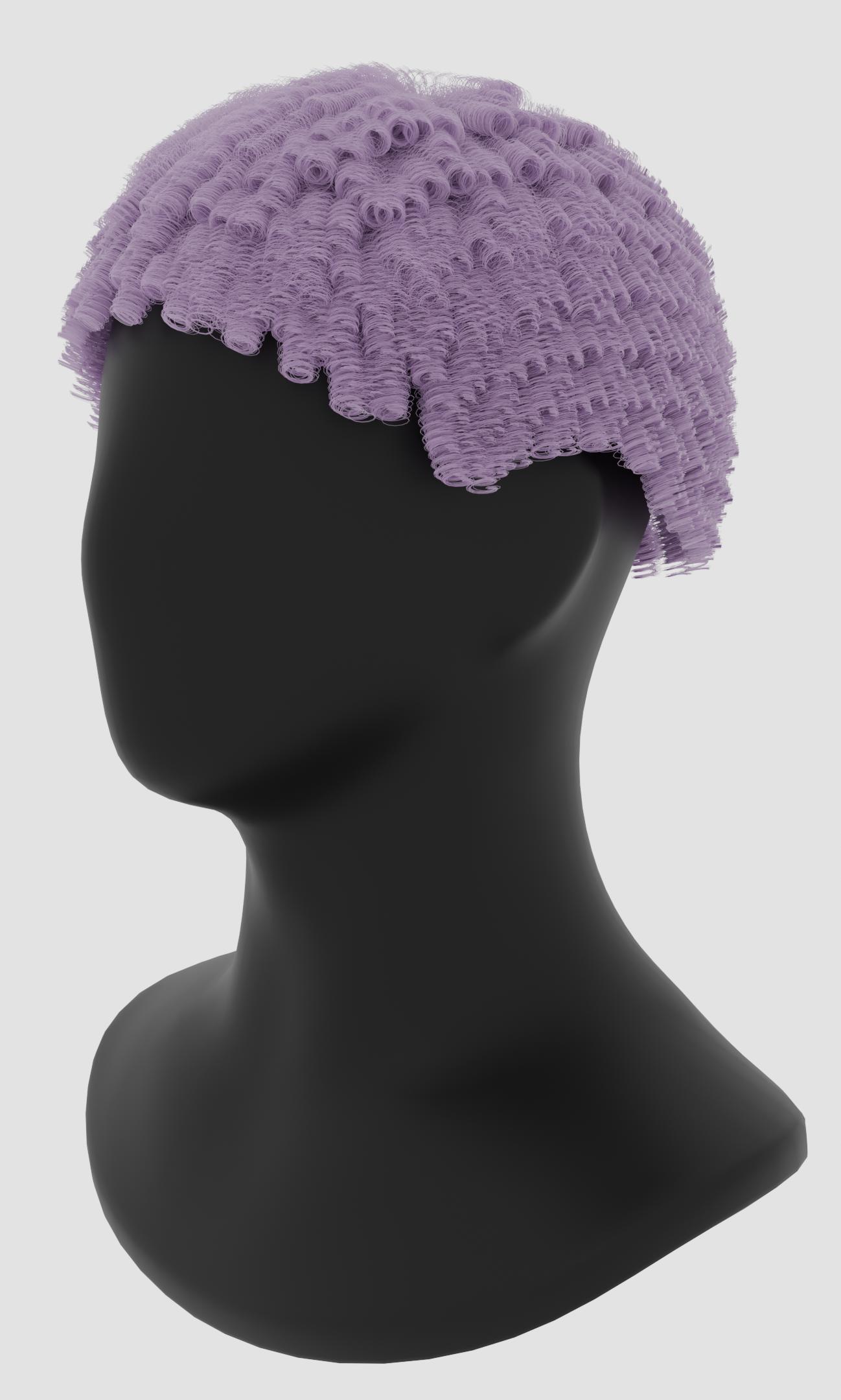}}
    \caption{More curly hair styles simulated using our wave-based hair model, showcasing its versatility in capturing diverse wavy and spiral hair dynamics.}
    \label{fig:more-hair-styles}
\end{figure}

We have applied our wave-based hair model to simulate a variety of curly hair styles, showcasing its versatility and effectiveness in capturing high-frequency wavy and spiral hair dynamics. As illustrated in \autoref{fig:teaser}, \ref{fig:more-hair-styles}, our model successfully represents diverse curly hair configurations, including tightly coiled curls, loose waves, and complex spiral patterns. Each hairstyle is constructed using discrete elements with independently parameterized wave characteristics, allowing for fine control over the appearance and behavior of individual hair strands. The simulations demonstrate the model's ability to accurately capture the intricate geometry of curly hair while efficiently handling dynamic deformations such as stretching, bending, and twisting. The results highlight the model's capacity to maintain physical realism and visual fidelity across a range of curly hair styles, making it a powerful tool for computer graphics and animation applications.

\subsection{Performance}
\label{sec:performance}

Table~\ref{tab:performance} summarizes the computational characteristics of our curly-element simulator across representative hairstyles and compares it with a Discrete Elastic Rod (DER) baseline~\cite{Bergou2008}. For each case, we report the problem size in terms of strands, elements, vertices, and degrees of freedom (DoFs), together with a per-frame timing breakdown of the total per-frame time (TPF, ms) into collision detection, Hessian assembly, and the preconditioned conjugate gradient (PCG) solve. Percentages indicate the share of TPF attributable to each stage.

The final columns present DER results at different discretization levels. DoFs-2 denotes a finer DER discretization with approximately $3-5\times$ more DoFs of our model, yielding comparable visual quality but incurring $3–9\times$ higher runtime, whereas DoFs-1 corresponds to a coarser discretization with roughly half the DoFs but substantially degraded results. This comparison demonstrates that conventional rod-based methods scale inefficiently with geometric resolution, while our wave-based model achieves high-fidelity curly hair dynamics at significantly lower computational cost.

\subsection{Comparisons}

\begin{figure}[t]
    \centering
    \subfigure[DER 8mm]{\includegraphics[width=0.32\linewidth]{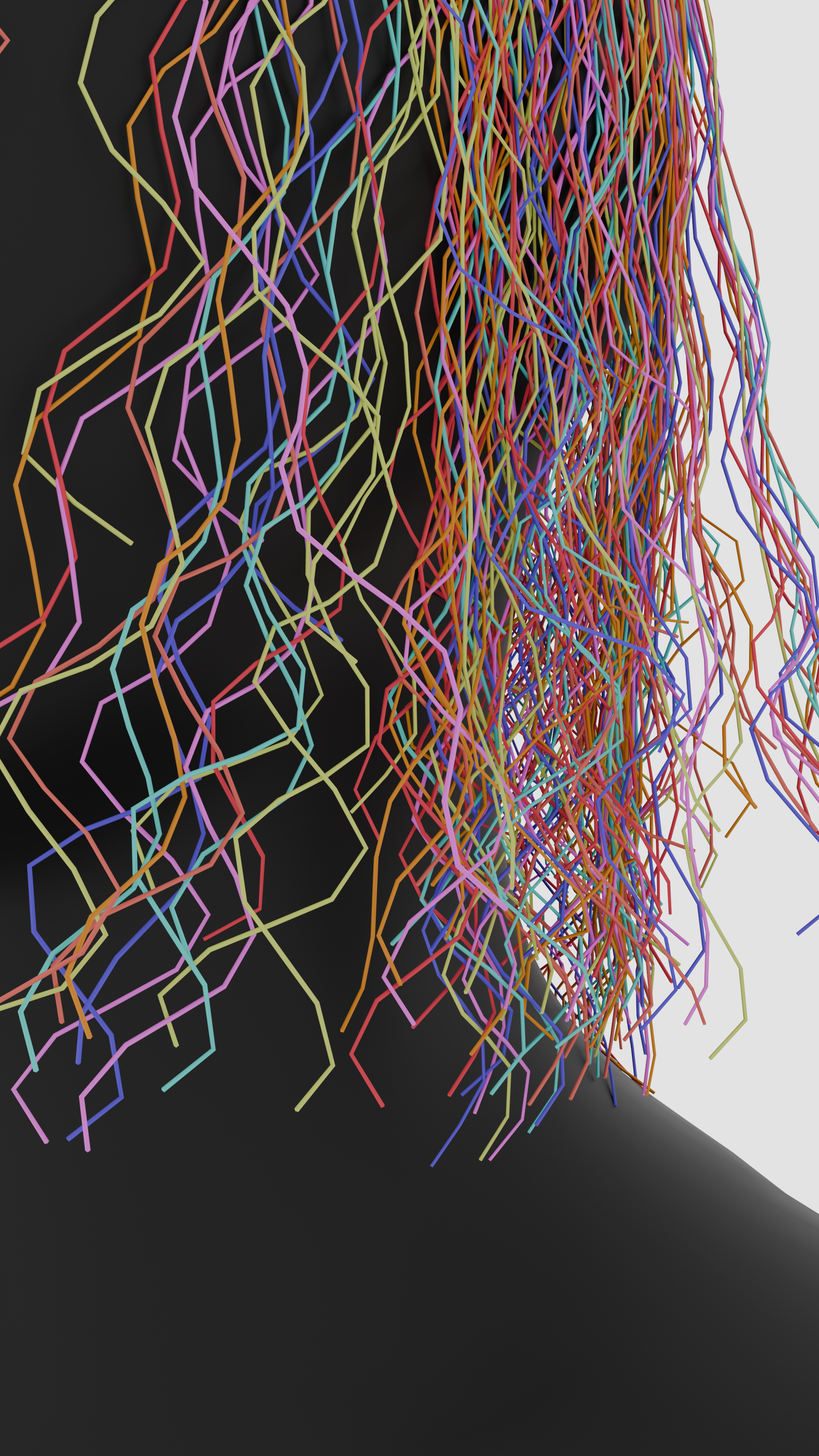}}
    \subfigure[DER 2mm]{\includegraphics[width=0.32\linewidth]{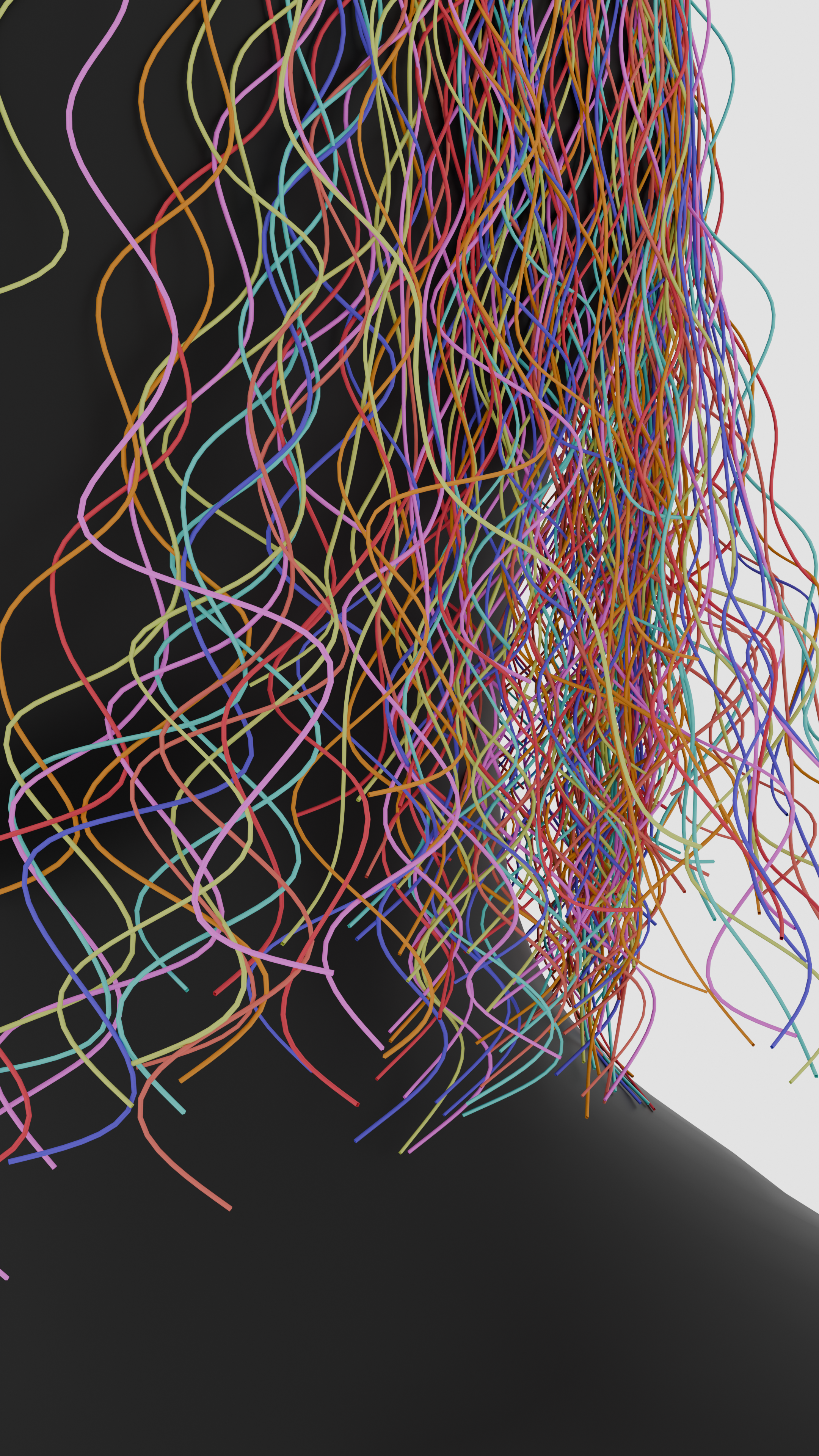}}
    \subfigure[Our Method]{\includegraphics[width=0.32\linewidth]{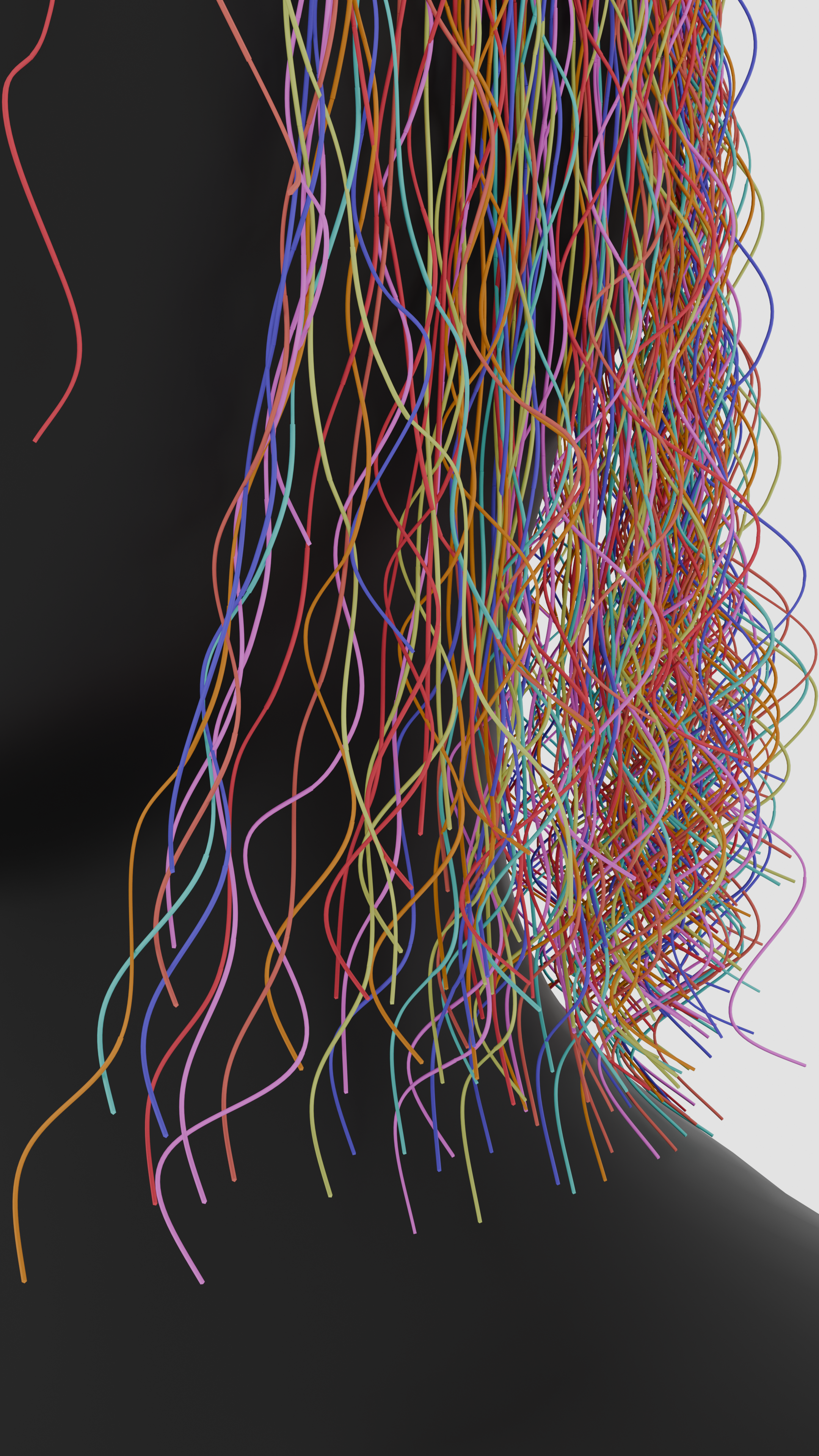}}
    \caption{Comparison to baseline method DER \cite{Bergou2008} at varying resolutions. Our wave-based model (c) captures high-frequency wavy details with significantly fewer degrees of freedom compared to DER at both coarse (a) and fine (b) discretizations.}
    \label{fig:comparison_der}
\end{figure}

As shown in \autoref{fig:comparison_der} and \autoref{tab:performance}, we compare our wave-based hair model with the Discrete Elastic Rod (DER) method~\cite{Bergou2008} for simulating a wavy hair strand. DER requires fine discretization ($2mm$ elements) to capture high-frequency details, incurring substantial computational cost. In contrast, our model achieves comparable visual fidelity using far fewer curly elements, significantly reducing degrees of freedom and simulation time. Coarser DER discretizations (e.g., $8mm$ elements) match our speed but introduce noticeable artifacts. These results demonstrate the efficiency of our approach in simulating complex curly hair dynamics while preserving detail and realism.

\subsection{Conclusion}

% To conclude, we have presented two novel curly elements, wavy element and spiral element for curly hair simulation that efficiently captures high-frequency wavy and spiral hair dynamics through a continuous representation of microscopic details. Our approach leverages a discrete finite element framework, enabling accurate simulation of complex hair behaviors while significantly reducing computational costs compared to traditional methods. The proposed curvature energy splitting and angle approximation techniques further enhance the model's physical realism and computational efficiency. Our results demonstrate the effectiveness of our model in various simulation scenarios, showcasing its potential for applications in computer graphics and animation.
In summary, we have introduced two novel curly elements—wavy and spiral—for simulating curly hair, enabling efficient capture of high-frequency hair dynamics via a continuous representation of microscopic details. Leveraging a discrete finite-element framework, our approach accurately models complex hair behaviors while substantially reducing computational cost compared to conventional methods. The proposed curvature energy splitting and angle-approximation techniques further enhance physical fidelity and efficiency. Experimental results demonstrate the model’s effectiveness across diverse scenarios, highlighting its potential for computer graphics and animation applications.

\section{Limitations and Future work}

% While our wave-based hair model offers significant advantages in simulating high-frequency wavy and spiral hair dynamics, it is not without limitations. One primary limitation is the assumption of a fixed number of wrinkles or helical turns during deformation, which may not accurately capture scenarios involving extreme stretching or compression where new wrinkles could form or existing ones could disappear. Additionally, our model currently focuses on bending and torsional energies, potentially overlooking other mechanical effects such as shear forces that may play a role in certain hair dynamics. Furthermore, the discretization approach, while efficient, may introduce artifacts in scenarios involving complex interactions between multiple hair strands, particularly in dense hair configurations. Finally, while our collision handling mechanism is effective for moderate hair densities, it may struggle with very dense hair or intricate interactions with external objects, leading to potential inaccuracies or instabilities. 

While our wave-based hair model effectively captures high-frequency wavy and spiral dynamics, it has several limitations. It assumes a fixed number of wrinkles or helical turns, limiting accuracy under extreme stretching or compression where wrinkles may form or vanish. The model primarily accounts for bending and torsional energies, potentially neglecting other mechanical effects such as shear. Although the discretization is efficient, it may introduce artifacts in dense or highly interacting hair strands. Finally, while the collision handling is robust for moderate densities, very dense hair or complex interactions with external objects may cause inaccuracies or instabilities.

For future work, we plan to extend our model to incorporate additional mechanical effects, including shear forces and more complex inter-strand interactions. We also aim to enhance collision handling to robustly support dense hair configurations and environmental interactions. Additionally, we will explore alternative element types and adaptive discretization strategies to dynamically adjust detail levels, further improving computational efficiency while maintaining simulation accuracy.

\bibliographystyle{ACM-Reference-Format}
\bibliography{ref_paper}

@inproceedings{Selle2008,
author = {Selle, Andrew and Lentine, Michael and Fedkiw, Ronald},
title = {A mass spring model for hair simulation},
year = {2008},
isbn = {9781450301121},
publisher = {Association for Computing Machinery},
address = {New York, NY, USA},
url = {https://doi.org/10.1145/1399504.1360663},
doi = {10.1145/1399504.1360663},
booktitle = {ACM SIGGRAPH 2008 Papers},
articleno = {64},
numpages = {11},
location = {Los Angeles, California},
series = {SIGGRAPH '08}
}

@inproceedings{Iben2013,
author = {Iben, Hayley and Meyer, Mark and Petrovic, Lena and Soares, Olivier and Anderson, John and Witkin, Andrew},
title = {Artistic simulation of curly hair},
year = {2013},
isbn = {9781450321327},
publisher = {Association for Computing Machinery},
address = {New York, NY, USA},
url = {https://doi.org/10.1145/2485895.2485913},
doi = {10.1145/2485895.2485913},
booktitle = {Proceedings of the 12th ACM SIGGRAPH/Eurographics Symposium on Computer Animation},
pages = {63–71},
numpages = {9},
location = {Anaheim, California},
series = {SCA '13}
}

@article{Yuksel2009,
author = {Yuksel, Cem and Schaefer, Scott and Keyser, John},
title = {Hair meshes},
year = {2009},
issue_date = {December 2009},
publisher = {Association for Computing Machinery},
address = {New York, NY, USA},
volume = {28},
number = {5},
issn = {0730-0301},
url = {https://doi.org/10.1145/1618452.1618512},
doi = {10.1145/1618452.1618512},
journal = {ACM Trans. Graph.},
month = dec,
pages = {1–7},
numpages = {7}
}

@inproceedings{Wu2016,
author = {Wu, Kui and Yuksel, Cem},
title = {Real-time hair mesh simulation},
year = {2016},
isbn = {9781450340434},
publisher = {Association for Computing Machinery},
address = {New York, NY, USA},
url = {https://doi.org/10.1145/2856400.2856412},
doi = {10.1145/2856400.2856412},
booktitle = {Proceedings of the 20th ACM SIGGRAPH Symposium on Interactive 3D Graphics and Games},
pages = {59–64},
numpages = {6},
location = {Redmond, Washington},
series = {I3D '16}
}

@inproceedings{Spillmann2007,
author = {Spillmann, J. and Teschner, M.},
title = {CoRdE: Cosserat rod elements for the dynamic simulation of one-dimensional elastic objects},
year = {2007},
isbn = {9781595936240},
publisher = {Eurographics Association},
address = {Goslar, DEU},
booktitle = {Proceedings of the 2007 ACM SIGGRAPH/Eurographics Symposium on Computer Animation},
pages = {63–72},
numpages = {10},
location = {San Diego, California},
series = {SCA '07}
}

@article{Bergou2008,
author = {Bergou, Mikl\'{o}s and Wardetzky, Max and Robinson, Stephen and Audoly, Basile and Grinspun, Eitan},
title = {Discrete elastic rods},
year = {2008},
issue_date = {August 2008},
publisher = {Association for Computing Machinery},
address = {New York, NY, USA},
volume = {27},
number = {3},
issn = {0730-0301},
url = {https://doi.org/10.1145/1360612.1360662},
doi = {10.1145/1360612.1360662},
journal = {ACM Trans. Graph.},
month = aug,
pages = {1–12},
numpages = {12}
}

@article{Bergou2010,
author = {Bergou, Mikl\'{o}s and Audoly, Basile and Vouga, Etienne and Wardetzky, Max and Grinspun, Eitan},
title = {Discrete viscous threads},
year = {2010},
issue_date = {July 2010},
publisher = {Association for Computing Machinery},
address = {New York, NY, USA},
volume = {29},
number = {4},
issn = {0730-0301},
url = {https://doi.org/10.1145/1778765.1778853},
doi = {10.1145/1778765.1778853},
journal = {ACM Trans. Graph.},
month = jul,
articleno = {116},
numpages = {10}
}

@article{Shi2023,
author = {Shi, Alvin and Wu, Haomiao and Parr, Jarred and Darke, A. M. and Kim, Theodore},
title = {Lifted Curls: A Model for Tightly Coiled Hair Simulation},
year = {2023},
issue_date = {August 2023},
publisher = {Association for Computing Machinery},
address = {New York, NY, USA},
volume = {6},
number = {3},
url = {https://doi.org/10.1145/3606920},
doi = {10.1145/3606920},
journal = {Proc. ACM Comput. Graph. Interact. Tech.},
month = aug,
articleno = {42},
numpages = {19}
}

@inproceedings{Wu2024,
author = {Wu, Haomiao and Shi, Alvin and Darke, A.M. and Kim, Theodore},
title = {Curly-Cue: Geometric Methods for Highly Coiled Hair},
year = {2024},
isbn = {9798400711312},
publisher = {Association for Computing Machinery},
address = {New York, NY, USA},
url = {https://doi.org/10.1145/3680528.3687641},
doi = {10.1145/3680528.3687641},
booktitle = {SIGGRAPH Asia 2024 Conference Papers},
articleno = {112},
numpages = {11},
location = {Tokyo, Japan},
series = {SA '24}
}

@inproceedings{Darke2024,
author = {Darke, A. M. and Olander, Isaac and Kim, Theodore},
title = {More Than Killmonger Locs: A Style Guide for Black Hair (in Computer Graphics)},
year = {2024},
isbn = {9798400706837},
publisher = {Association for Computing Machinery},
address = {New York, NY, USA},
url = {https://doi.org/10.1145/3664475.3664535},
doi = {10.1145/3664475.3664535},
booktitle = {ACM SIGGRAPH 2024 Courses},
articleno = {18},
numpages = {251},
location = {Denver, CO, USA},
series = {SIGGRAPH '24 Courses}
}

@inproceedings{Tariq2008,
author = {Tariq, Sarah and Bavoil, Louis},
title = {Real time hair simulation and rendering on the GPU},
year = {2008},
isbn = {9781605583433},
publisher = {Association for Computing Machinery},
address = {New York, NY, USA},
url = {https://doi.org/10.1145/1401032.1401080},
doi = {10.1145/1401032.1401080},
booktitle = {ACM SIGGRAPH 2008 Talks},
articleno = {37},
numpages = {1},
location = {Los Angeles, California},
series = {SIGGRAPH '08}
}

@inproceedings{Somasundaram2015,
author = {Somasundaram, Arunachalam},
title = {Dynamically controlling hair interpolation},
year = {2015},
isbn = {9781450336369},
publisher = {Association for Computing Machinery},
address = {New York, NY, USA},
url = {https://doi.org/10.1145/2775280.2792541},
doi = {10.1145/2775280.2792541},
booktitle = {ACM SIGGRAPH 2015 Talks},
articleno = {36},
numpages = {1},
location = {Los Angeles, California},
series = {SIGGRAPH '15}
}

@ARTICLE{Lyu2022,
  author={Lyu, Qing and Chai, Menglei and Chen, Xiang and Zhou, Kun},
  journal={IEEE Transactions on Visualization and Computer Graphics}, 
  title={Real-Time Hair Simulation With Neural Interpolation}, 
  year={2022},
  volume={28},
  number={4},
  pages={1894-1905},
  doi={10.1109/TVCG.2020.3029823}}

@article{Hsu2024,
author = {Hsu, Jerry and Wang, Tongtong and Pan, Zherong and Gao, Xifeng and Yuksel, Cem and Wu, Kui},
title = {Real-time Physically Guided Hair Interpolation},
year = {2024},
issue_date = {July 2024},
publisher = {Association for Computing Machinery},
address = {New York, NY, USA},
volume = {43},
number = {4},
issn = {0730-0301},
url = {https://doi.org/10.1145/3658176},
doi = {10.1145/3658176},
journal = {ACM Trans. Graph.},
month = jul,
articleno = {95},
numpages = {11}
}

@inproceedings{Umetani2015,
author = {Umetani, Nobuyuki and Schmidt, Ryan and Stam, Jos},
title = {Position-based elastic rods},
year = {2015},
publisher = {Eurographics Association},
address = {Goslar, DEU},
booktitle = {Proceedings of the ACM SIGGRAPH/Eurographics Symposium on Computer Animation},
pages = {21–30},
numpages = {10},
location = {Copenhagen, Denmark},
series = {SCA '14}
}

@inproceedings{Hsu2025,
author = {Hsu, Jerry and Wang, Tongtong and Wu, Kui and Yuksel, Cem},
title = {Stable Cosserat Rods},
year = {2025},
isbn = {9798400715402},
publisher = {Association for Computing Machinery},
address = {New York, NY, USA},
url = {https://doi.org/10.1145/3721238.3730618},
doi = {10.1145/3721238.3730618},
booktitle = {Proceedings of the Special Interest Group on Computer Graphics and Interactive Techniques Conference Conference Papers},
articleno = {75},
numpages = {10},
location = {
},
series = {SIGGRAPH Conference Papers '25}
}

@inproceedings{Daviet2023hair,
author = {Daviet, Gilles},
title = {Interactive Hair Simulation on the GPU using ADMM},
year = {2023},
isbn = {9798400701597},
publisher = {Association for Computing Machinery},
address = {New York, NY, USA},
url = {https://doi.org/10.1145/3588432.3591551},
doi = {10.1145/3588432.3591551},
booktitle = {ACM SIGGRAPH 2023 Conference Proceedings},
articleno = {24},
numpages = {11},
location = {Los Angeles, CA, USA},
series = {SIGGRAPH '23}
}

@article{ward2007survey,
  title={A survey on hair modeling: Styling, simulation, and rendering},
  author={Ward, Kelly and Bertails, Florence and Kim, Tae-Yong and Marschner, Stephen R and Cani, Marie-Paule and Lin, Ming C},
  journal={IEEE transactions on visualization and computer graphics},
  volume={13},
  number={2},
  pages={213--234},
  year={2007},
  publisher={IEEE}
}

@inproceedings{Bertails2006superhelices,
author = {Bertails, Florence and Audoly, Basile and Cani, Marie-Paule and Querleux, Bernard and Leroy, Fr\'{e}d\'{e}ric and L\'{e}v\^{e}que, Jean-Luc},
title = {Super-helices for predicting the dynamics of natural hair},
year = {2006},
isbn = {1595933646},
publisher = {Association for Computing Machinery},
address = {New York, NY, USA},
url = {https://doi.org/10.1145/1179352.1142012},
doi = {10.1145/1179352.1142012},
booktitle = {ACM SIGGRAPH 2006 Papers},
pages = {1180–1187},
numpages = {8},
location = {Boston, Massachusetts},
series = {SIGGRAPH '06}
}

@inproceedings{hadap2001modeling,
  title={Modeling dynamic hair as a continuum},
  author={Hadap, Sunil and Magnenat-Thalmann, Nadia},
  booktitle={Computer Graphics Forum},
  volume={20},
  number={3},
  pages={329--338},
  year={2001},
  organization={Wiley Online Library}
}

@inproceedings{bando2003animating,
  title={Animating hair with loosely connected particles},
  author={Bando, Yosuke and Chen, Bing-Yu and Nishita, Tomoyuki},
  booktitle={Computer Graphics Forum},
  volume={22},
  number={3},
  pages={411--418},
  year={2003},
  organization={Wiley Online Library}
}

@inproceedings{Daviet2011hair,
author = {Daviet, Gilles and Bertails-Descoubes, Florence and Boissieux, Laurence},
title = {A hybrid iterative solver for robustly capturing coulomb friction in hair dynamics},
year = {2011},
isbn = {9781450308076},
publisher = {Association for Computing Machinery},
address = {New York, NY, USA},
url = {https://doi.org/10.1145/2024156.2024173},
doi = {10.1145/2024156.2024173},
booktitle = {Proceedings of the 2011 SIGGRAPH Asia Conference},
articleno = {139},
numpages = {12},
location = {Hong Kong, China},
series = {SA '11}
}

@article{Hu2014robusthair,
author = {Hu, Liwen and Ma, Chongyang and Luo, Linjie and Li, Hao},
title = {Robust hair capture using simulated examples},
year = {2014},
issue_date = {July 2014},
publisher = {Association for Computing Machinery},
address = {New York, NY, USA},
volume = {33},
number = {4},
issn = {0730-0301},
url = {https://doi.org/10.1145/2601097.2601194},
doi = {10.1145/2601097.2601194},
journal = {ACM Trans. Graph.},
month = jul,
articleno = {126},
numpages = {10}
}

@article{Dominik2015,

title = {A physically based approach to the accurate simulation of stiff fibers and stiff fiber meshes},

journal = {Computers \& Graphics},

volume = {53},

pages = {136-146},

year = {2015},

issn = {0097-8493},

author = {Dominik L. Michels and J. Paul T. Mueller and Gerrit A. Sobottka}

}

@article{Fei2021,

author = {Fei, Yun (Raymond) and Guo, Qi and Wu, Rundong and Huang, Li and Gao, Ming},

title = {Revisiting integration in the material point method: a scheme for easier separation and less dissipation},

year = {2021},

issue_date = {August 2021},

publisher = {Association for Computing Machinery},

address = {New York, NY, USA},

volume = {40},

number = {4},

issn = {0730-0301},

journal = {ACM Trans. Graph.},

month = jul,

articleno = {109},

numpages = {16}

}

@article{Hsu2023,

author = {Hsu, Jerry and Wang, Tongtong and Pan, Zherong and Gao, Xifeng and Yuksel, Cem and Wu, Kui},

title = {Sag-Free Initialization for Strand-Based Hybrid Hair Simulation},

year = {2023},

issue_date = {August 2023},

publisher = {Association for Computing Machinery},

address = {New York, NY, USA},

volume = {42},

number = {4},

issn = {0730-0301},

journal = {ACM Trans. Graph.},

month = jul,

articleno = {74},

numpages = {14}

}

@inproceedings{Koh2001,

author = {Koh, Chuan Koon and Huang, Zhiyong},

title = {A simple Physics model to animate human hair modeled in 2D strips in real time},

year = {2001},

isbn = {3211837116},

publisher = {Springer-Verlag},

address = {Berlin, Heidelberg},

booktitle = {Proceedings of the Eurographic Workshop on Computer Animation and Simulation},

pages = {127–138},

numpages = {12},

location = {Manchester, UK}

}

@inproceedings{Volino2004,

author = {Volino, Pascal and Magnenat-Thalmann, Nadia},

title = {Animating complex hairstyles in real-time},

year = {2004},

isbn = {1581139071},

publisher = {Association for Computing Machinery},

address = {New York, NY, USA},

url = {https://doi.org/10.1145/1077534.1077544},

doi = {10.1145/1077534.1077544},

booktitle = {Proceedings of the ACM Symposium on Virtual Reality Software and Technology},

pages = {41–48},

numpages = {8},

location = {Hong Kong},

series = {VRST '04}

}

@article{Han2019,

author = {Han, Xuchen and Gast, Theodore F. and Guo, Qi and Wang, Stephanie and Jiang, Chenfanfu and Teran, Joseph},

title = {A Hybrid Material Point Method for Frictional Contact with Diverse Materials},

year = {2019},

issue_date = {July 2019},

publisher = {Association for Computing Machinery},

address = {New York, NY, USA},

volume = {2},

number = {2},

url = {https://doi.org/10.1145/3340258},

doi = {10.1145/3340258},

journal = {Proc. ACM Comput. Graph. Interact. Tech.},

month = jul,

articleno = {17},

numpages = {24}

}

\clearpage

% \appendix

\end{document}